\documentclass[journal]{IEEEtran}
\usepackage{amsbsy}
\usepackage{floatflt} 

\usepackage{amsmath}
\usepackage{amssymb}
\usepackage{times}
\usepackage{graphics}
\usepackage{graphicx}
\usepackage{xspace}
\usepackage{paralist} 
\usepackage{setspace} 
\usepackage{xypic}
\xyoption{curve}
\usepackage{latexsym}
\usepackage{theorem}
\usepackage{ifthen}
\usepackage{caption}
\usepackage{subfigure}
\usepackage{booktabs}
\usepackage{algorithm}
\usepackage{algorithmic}
\usepackage{color}

\ifCLASSINFOpdf
\else
\fi

{\theoremheaderfont{\it} \theorembodyfont{\rmfamily}
\newtheorem{theorem}{Theorem}
\newtheorem{lemma}[theorem]{Lemma}

}

\newcounter{brojac}

{\theoremheaderfont{\it} \theorembodyfont{\rmfamily}
\newtheorem{assumption}[brojac]{Assumption}
}

\newcounter{brojac2}

{\theoremheaderfont{\it} \theorembodyfont{\rmfamily}
\newtheorem{remark}[brojac2]{Remark}
}

\begin{document}
\title{A Unification and Generalization of Exact Distributed First Order Methods}

\author{Du$\check{\mbox{s}}$an Jakoveti\'c
\thanks{D. Jakoveti\'c is with the Department of Mathematics and Informatics, Faculty of Sciences, University of Novi Sad, Novi Sad, Serbia. The research is supported by Ministry of Education, Science and Technological Development, Republic of Serbia, grant no. 174030. Author's e-mail: djakovet@uns.ac.rs.}}

%
%


\maketitle

\begin{abstract}
Recently, there has been significant progress in the development of distributed
first order methods. (At least) two different types of methods,
designed from very different perspectives, have been proposed
that achieve both \emph{exact} and \emph{linear} convergence when a constant step size is used --
a favorable feature that was not achievable by most prior methods.
In this paper, we unify, generalize, and improve convergence speed of these exact distributed
first order methods. We first carry out a novel unifying
analysis that sheds light on how the
different existing methods compare. The analysis reveals that
a major difference between the methods is
on how a past dual gradient of an associated augmented Lagrangian dual function is weighted.
 We then capitalize on the insights from the analysis
to derive a novel method -- with a tuned past gradient weighting --
that improves upon the existing methods.
  We establish for the proposed generalized method
 global R-linear convergence rate under strongly convex costs with Lipschitz continuous gradients.

\end{abstract}


\maketitle \thispagestyle{empty} \maketitle
%
%
%
%
\vspace{-4mm}

\section{Introduction}

\textbf{Context and motivation}. Distributed optimization methods for solving
convex optimization problems over networks, e.g.,~\cite{nedic_T-AC}--\nocite{SayedConf,SayedEstimation,SayedMagazineDiffusion,duchi,asu-random,arxivVersion}\cite{randomNesterov}, have gained a significant renewed and growing interest over the
last decade, motivated by
various applications, ranging from
inference problems in sensor networks, e.g.,~\cite{SoummyaEst,Rabbat},
to distributed learning, e.g.,~\cite{BoydADMoM}, to distributed control problems, e.g.,~\cite{BulloBook}. 
%
%

Recently, there have been significant advances
in the context of \emph{distributed first order methods}.
 A distinctive feature of the novel methods, proposed and analyzed in~\cite{WotaoYinExtra}--\nocite{Harnessing,SayedExact1,SayedExact2,LinearRateAL,SmallGainNedicUncoord,SmallGainNedicTimeVar,SmallGainKin1,SmallGainKin2,HarnessingNesterov,UsmanDextra,ExtraPush,PGextra}\cite{FrensisBachDistributedOpt},   is that,
unlike most of prior distributed first order algorithms, e.g.,~\cite{nedic_T-AC,nedic_T-AC-private,duchi,arxivVersion}, they converge to the exact solution
 even when a constant (non-diminishing) step size is used.
This property allows the methods to achieve global linear convergence rates, when
the nodes' local costs are strongly convex and have Lipschitz
continuous gradients. Among
the exact distributed first order methods,
 (at least) two different types of methods have been proposed, each
 designed through a very different methodology.
The method in~\cite{WotaoYinExtra}, dubbed Extra, see also~\cite{PGextra,UsmanDextra}, modifies
the update of the standard
distributed gradient method, e.g.,~\cite{nedic_T-AC},
by introducing two different sets of weighting coefficients (weight matrices)
for any two consecutive iterations of the algorithm, as opposed to a
single weight matrix with the standard method~\cite{nedic_T-AC}.
The second type of methods,~\cite{Harnessing,SmallGainNedicTimeVar,SmallGainNedicUncoord,SmallGainKin1,SmallGainKin2},
 replaces the nodes' local gradient with
a ``tracked'' value of the network-wide
average of the nodes' local gradients.
  The method in \cite{Harnessing}
 has been proved to be convergent under time-varying networks, e.g.,~\cite{SmallGainNedicTimeVar}, node-varying
 step sizes, e.g.,\cite{SmallGainNedicUncoord}, and Nesterov acceleration~\cite{HarnessingNesterov},
 while such studies have not been provided to date for the method in~\cite{WotaoYinExtra}.
  Both references~\cite{WotaoYinExtra} and~\cite{Harnessing}
establish global linear convergence
rates of the exact distributed first order methods
that they study.
Further, reference~\cite{MokhtariDoubly}
shows that Extra is equivalent to a primal-dual gradient-like method (see, e.g.,~\cite{SaddlePointNedicAsu,SaddlePoint2,SaddlePointUzawa}, for primal-dual (sub)gradient methods)  applied on the augmented
Lagrangian dual problem
of a reformulation of the
original problem of interest.
 Reference~\cite{SmallGainNedicTimeVar}
demonstrates that the method therein, equivalent to the method in~\cite{Harnessing},
 can be put in the form of Extra, with a
 specific choice of the two Extra's weight matrices,
 and it also provides
 a primal-dual interpretation of the exact method therein.
 (A more detailed review of works~\cite{WotaoYinExtra}--\nocite{Harnessing,SayedExact1,SayedExact2,LinearRateAL,SmallGainNedicUncoord,SmallGainNedicTimeVar,SmallGainKin1,SmallGainKin2,HarnessingNesterov,UsmanDextra,ExtraPush,PGextra}\cite{FrensisBachDistributedOpt}   is provided further ahead.)

\textbf{Contributions}. The main contributions of this paper
are to unify, generalize, and improve convergence speed of exact distributed first order methods.
%
  First, with both the methods in \cite{WotaoYinExtra}
  and \cite{Harnessing},  we provide an
  augmented Lagrangian-based analysis. that
  gives for both methods a characterization
 of the (joint) evolution of primal and dual
 errors along iterations. This characterization reveals
 the effect of the difference between the two methods on their
 performance and sheds light
 on understanding of how the two methods compare under various problem model scenarios.
  We further provide a generalized
 method, parameterized with an additional, easy-to-tune network-wide
 weight matrix, that determines the weighting of the
 term that corresponds to a past dual gradient of the
 associated augmented Lagrangian function.
 The generalized method subsumes the two known
 methods in~\cite{WotaoYinExtra} and~\cite{Harnessing} upon setting a weighting
 parameter matrix to appropriate
 specific values. We
 describe how to set the weighting
 parameter matrix to improve upon both
 existing methods.
 The proposed parameter matrix tuning is based
 on an insight from our analysis
 on how the negative effect of the primal error
 on the dual error can be reduced (compensated)
 through the introduced parameter matrix.


The paper carries out convergence rate
analysis of the proposed generalized method when nodes' local costs are strongly convex,
have Lipschitz continuous gradient, and the
underlying network is static. With the proposed generalized method, we establish a global
R-linear convergence rate, when the algorithm's step size is appropriately set.
 Numerical examples confirm the insights from our
analysis on the mutual comparison
of the methods in~\cite{WotaoYinExtra}
and~\cite{Harnessing},
as well as the improvements of the proposed generalized method
over the two existing ones.

\textbf{Brief literature review}. Distributed computation and optimization
has been studied for a long time, e.g.,~\cite{Tsitsiklis_Distr_Opt}.
More recently, reference~\cite{nedic_T-AC}
proposes a distributed first order (subgradient)
method for unconstrained problems, allowing for
possibly non-differentiable, convex nodes' local costs and carries out its convergence
and convergence rate analysis for
deterministically time varying networks.
 A distributed projected subgradient method
 for constrained problems and possibly non-differentiable local costs has been proposed and analyzed in~\cite{nedic_T-AC-private}.
 References~\cite{arxivVersion,LinearRateAL,WotaoYinExtra,Harnessing,SayedExact1,SayedExact2,SmallGainNedicUncoord,SmallGainNedicTimeVar,SmallGainKin1,SmallGainKin2,HarnessingNesterov}
  study unconstrained distributed optimization
  under more structured local costs. In~\cite{arxivVersion},
  distributed gradient methods
  with an acceleration based
  on the Nesterov (centralized) gradient method~\cite{Nesterov-Gradient}
   have been proposed and analyzed under differentiable local costs
   with Lipschitz continuous and bounded gradients.
   The methods in~\cite{nedic_T-AC,nedic_T-AC-private,arxivVersion,TSPvarNoWorkNodes}
      converge to the exact solution only when a diminishing step-size is used; when a constant step size is used
      they converge to a solution neighborhood.
    References~\cite{MartinezPrimalDual,NedicPrimalDual}
    propose different types of primal-dual methods,
    prove their convergence to the exact solution for a wide class of problems assuming diminishing
    step-sizes, and are not concerned with establishing the methods' convergence rates. The authors of~\cite{WangEliaControlApproachDopt}
    use insights from control theory to propose a gradient-like algorithm for which
    they prove exact convergence under certain conditions, while
    the paper is not concerned with analyzing the method's convergence rate. Reference~\cite{LinearRateAL}
    proposes several variants of distributed AL
    methods that converge to the exact solution under a constant step size and establishes their
    linear convergence rates for twice
    continuously differentiable
    costs with bounded Hessian.

       References~\cite{WotaoYinExtra,Harnessing,SayedExact1,SayedExact2,SmallGainNedicUncoord,SmallGainNedicTimeVar,SmallGainKin1,SmallGainKin2,HarnessingNesterov}
   develop and/or analyze different variants of \emph{exact} distributed first order methods under various
   assumptions on the nodes' local costs, algorithm step sizes, and the underlying network.
    The papers~\cite{WotaoYinExtra,Harnessing}
    propose two different exact distributed first order methods and analyze their convergence rates, as
    already discussed above.
     References~\cite{SayedExact1,SayedExact2}
     develop exact methods based
     on diffusion algorithms (see, e.g.,~\cite{SayedEstimation}) and
     through a primal-dual type method on an associated AL function.
     Under twice differentiable local costs, each node's cost
      having Lipschitz continuous gradient, and at least one node's cost being strongly convex, the papers show linear convergence rates for the methods therein,
     allowing for different step sizes across nodes and for a wider range
     of step sizes and admissible weight matrices with respect to~\cite{WotaoYinExtra}.
     The papers~\cite{SmallGainNedicUncoord,SmallGainNedicTimeVar,SmallGainKin1,SmallGainKin2} consider strongly convex local costs with Lipschitz continuous gradients. Under this setting, reference~\cite{SmallGainNedicUncoord} establishes
 global linear convergence of the method studied
 therein when nodes utilize uncoordinated step sizes. The paper~\cite{SmallGainNedicTimeVar}
 establishes global linear convergence
 for time-varying networks.
 The authors of~\cite{SmallGainKin2}
  prove global linear rates under both time varying networks and
  uncoordinated step sizes, while the issue of
  uncoordinated step-sizes is previously
considered in~\cite{SmallGainKin1}. Finally,
  the paper~\cite{HarnessingNesterov}
  proposes an accelerated exact distributed first order
  method based on the Nesterov acceleration~\cite{Nesterov-Gradient}
  and establishes its convergence rates
  for local costs with Lipschitz continuous
  gradients, both in the presence and in absence
  of the strong convexity assumption.
  Under strongly convex local costs that have Lipschitz continuous gradients,
  the authors of~\cite{FrensisBachDistributedOpt}
  develop optimal distributed methods,
  where the optimality is in terms of the number of oracle calls of a therein
  appropriately defined oracle. However,
  their method is of a different type
  than~\cite{SmallGainNedicUncoord,SmallGainNedicTimeVar,SmallGainKin1,SmallGainKin2,HarnessingNesterov}
  and is different from the method proposed here. Namely, the method in~\cite{FrensisBachDistributedOpt}
  requires evaluation of
  Fenchel conjugates of the nodes'
  local costs at each iteration, and hence it in general has a much
  larger computational cost per iteration than the methods in~\cite{SmallGainNedicUncoord,SmallGainNedicTimeVar,SmallGainKin1,SmallGainKin2,HarnessingNesterov}
   and the method proposed in this paper.
    To the best of our knowledge,
   communication and
computation optimality of distributed first
   order methods that
   do not involve Fenchel conjugates has not been studied to date.



\textbf{Paper organization}.
 The next paragraph introduces notation. Section~{II}
  explains the model that we assume and
   reviews existing distributed first order methods.
  Section~{III} presents the proposed algorithm and relates
  it with the existing methods. Section~{IV}
  establishes global R-linear convergence rate of the proposed method, while
  Section~V provides simulation examples. Finally, we conclude in Section~{VI}.
    Certain auxiliary proofs are provided in the Appendix.

\textbf{Notation}. We denote by: $\mathbb R$ the set of real numbers; ${\mathbb R}^d$ the $d$-dimensional
Euclidean real coordinate space; $A_{ij}$ the entry in the $i$-th row and $j$-th column of a matrix $A$;
$A^\top$ the transpose of a matrix $A$;
$\otimes$ the Kronecker product of matrices; $I$, $0$, and $\mathbf{1}$, respectively, the identity matrix, the zero matrix, and the column vector with unit entries; $J$ the $N \times N$ matrix $J:=(1/N)\mathbf{1}\mathbf{1}^\top$;
 $A \succ  0 \,(A \succeq  0 )$ means that
 the symmetric matrix $A$ is positive definite (respectively, positive semi-definite);
  $\|\cdot\|=\|\cdot\|_2$ the Euclidean (respectively, spectral) norm of its vector (respectively, matrix) argument; $\lambda_i(\cdot)$ the $i$-th largest eigenvalue; $\mathrm{Diag}\left(a\right)$ the diagonal matrix with the diagonal equal to the vector $a$; $|\cdot|$ the cardinality of a set; $\nabla h(w)$ and $\nabla^2 h(w)$, respectively,
  the gradient and Hessian evaluated at $w$ of a function $h: {\mathbb R}^d \rightarrow {\mathbb R}$, $d \geq 1$.
 Finally, for two positive sequences $\eta_n$ and $\chi_n$, we have: $\eta_n = O(\chi_n)$ if
 $\limsup_{n \rightarrow \infty}\frac{\eta_n}{\chi_n}<\infty$; and $\eta_n = \Omega (\chi_n)$ if $\liminf_{n \rightarrow \infty}\frac{\eta_n}{\chi_n}>0$.

\vspace{-2mm}
\section{Model and preliminaries}

Subsection~{II}-A describes the optimization and network models
that we assume. Subsection~{II}-B reviews the standard
(inexact) distributed gradient method in~\cite{nedic_T-AC},
as well as the exact methods in~\cite{WotaoYinExtra} and~\cite{Harnessing},
 and it reviews the known equivalence (see~\cite{MokhtariDoubly}) of the
 method in~\cite{WotaoYinExtra} and a primal-dual gradient-like method.

\subsection{Optimization and network models}

We consider distributed optimization where
$N$ nodes in a connected network solve the following problem:
\begin{equation} \label{eqn-opt-prob-original}
\mathrm{min}_{x \in {\mathbb R}^d}\,\,f(x) := \sum_{i=1}^{N} f_i(x).
\end{equation}
Here, $f_i:\,{\mathbb R}^d \rightarrow \mathbb R$ is a convex
function known only by node~$i$.
 Throughout the paper, we impose the following assumption on the $f_i$'s.
\begin{assumption}
Each function $f_i:\,{\mathbb R}^d \rightarrow \mathbb R$, $i=1,...,N$, is strongly convex
with strong convexity parameter~$\mu$,
and it has Lipschitz continuous
gradient with Lipschitz constant~$L$,
where $L \geq \mu >0$. That is,
for all $i=1,...,N$, there holds:
\begin{eqnarray*}
&\,&f_i(y) \geq
f_i(x) + \nabla f_i(x)^\top (y-x) \\
&\,&\:\:\:\:\:\:\:\:\:\:\:\:\:\:\:\:\:\:\:\:\:\:\:\:+\, \frac{\mu}{2}\|x-y\|^2,\,x,y \in {\mathbb R}^d\\
&\,&\|\nabla f_i(x) - \nabla f_i(y)\|
\leq
L\,\|x-y\|,\,x,y \in {\mathbb R}^d.
\end{eqnarray*}
\end{assumption}
%
%
%
%
%
For a specific result (precisely, Lemma~\ref{lemma-recursion} ahead), we additionally assume the following.
\begin{assumption}
Each function $f_{i} : \mathbb{R}^d \rightarrow \mathbb{R}, \; i=1,...,N$, is twice continuously differentiable.
\end{assumption}
Under Assumptions 1 and 2, there holds for every $x \in \mathbb{R}^d$ that:
\[
\mu \,I \preceq \nabla^{2} f_{i}(x) \preceq L \,I.
\]
Further, under Assumption~1,
problem~\eqref{eqn-opt-prob-original}
is solvable and has the unique
solution~${x}^{\star} \in {\mathbb R}^d$.

Nodes $i=1,...,N$ constitute an undirected network
$ {\mathcal G} = ({\mathcal V},{E}) $,
where $\mathcal V$ is the set of nodes and
$E$ is the set of edges.
  The presence of edge $\{i,j\} \in E$
  means that the nodes~$i$ and $j$
   can directly exchange messages
   through a communication link.
 Further, let $\Omega_{i}$  be the set of all neighbors of a node $i$ (including~$i$).

\begin{assumption}
 The network  $ {\mathcal G} = ({\mathcal V},{E}) $ is connected, undirected and simple (no self-loops nor multiple links).
\end{assumption}

We associate with network $\mathcal G$
 a $N \times N$ symmetric, (doubly) stochastic weight matrix~$W$.
  Further, we let:
  $W_{ij}=W_{ji}>0$, if
 $\{i,j\} \in E$, $i \neq j$;
  $W_{ij}=W_{ji}=0$, if
 $\{i,j\} \notin E$, $i \neq j$;
 and $W_{ii} = 1 -\sum_{j \neq i} W_{ij}>0$,
 for all $i=1,...,N$.
   Denote by $\lambda_i$, $i=1,2,...,N$,
 the eigenvalues of $W$, ordered in a descending order;
 it can be shown that they obey
 $1=\lambda_1>\lambda_2 \geq ... \geq \lambda_N>-1.$

Throughout the paper, we consider several iterative distributed methods to
solve~\eqref{eqn-opt-prob-original}.
 An algorithm's iterations are indexed by $k=0,1,2,...$
 Further, we denote
by $x_i^{(k)} \in {\mathbb R}^d$ the estimate of the solution to~\eqref{eqn-opt-prob-original}
available to node $i$ at iteration~$k$.
To avoid notational clutter, we will keep the same notation $x_i^{(k)}$
across different methods, while it is clear from context which method
is in question. For compact notation,
we use $x^{(k)} = \left( (x_1^{(k)})^\top, (x_2^{(k)})^\top, ..., (x_N^{(k)})^\top \right)^\top
\in {\mathbb R}^{Nd}$ to denote the vector that stacks
the solution estimates  by all nodes at iteration~$k$. We
 use analogous notation for certain auxiliary sequences that
 the algorithms maintain (e.g., see ahead $s_i^{(k)}$ and $s^{(k)}$ with~\eqref{eqn-harnessing}.)
 Again, to simplify notation, with all methods to be considered we
 will assume equal initialization across all nodes, i.e.,
 we let $x_1^{(0)}=x_2^{(0)}=...=x_N^{(0)}$; e.g.,
 nodes can set $x_i^{(0)}=0$, for all~$i$.

For future reference, we introduce the following
 quantities. We let
the $(N d) \times (N d)$ matrix $\mathcal{W} = W \otimes I$,
where $I$ is the $d \times d $ identity matrix. That is,
$\mathcal{W}$  is a
$N \times N$ block matrix with
$d \times d$ blocks, such that
the block at the $(i,j)$-th position
equals $W_{ij} \,I$.\footnote{We will frequently
work with Kronecker products of types $A \otimes B$,
$a \otimes B$, and $a \otimes b$, where $A$ is an $N \times N$ matrix,
$a$ -- $N \times 1$ vector, $B$ -- $d \times d$ matrix,
and $b$ -- $d \times 1$ vector. In other words,
the Kronecker products throughout always appear
with the first argument of dimension either $N \times N$ or
$N \times 1$, and the second argument either $d \times d$ or $d \times 1$.
As the capital letters denote matrices and the lower case letters denote vectors,
 the dimensions of the Kronecker products arguments will be clear.}
 Note that the $(i,j)$-th and $(j,i)$-th block
 of $\mathcal{W}$ equal to zero
 if $\{i,j\} \notin E$, $i \neq j$.
 In other words, we say that
$\mathcal{W}$ respects
the sparsity pattern of the underlying graph~$\mathcal{G}$. Further, recall that
$J=\frac{1}{N}\mathbf{1}\mathbf{1}^\top$ is the
ideal consensus matrix,\footnote{The consensus
algorithm, e.g.,~\cite{dimakiskarmourarabbatscaglione},
computes the global average of nodes' local quantities
through a linear system iteration with
a stochastic system matrix $W$.
When $W=J$, consensus converges in a single iteration,
hence we name $J$ the ideal consensus matrix.
Clearly, $J$ is not realizable over a
generic graph as it is not
sparse, but it is usually
desirable to have $W$ as close as possible to
$J$ in an appropriate sense,
given the constraints on the network sparsity.}
and let
$\mathcal{J}=J \otimes I$. Denote
by $\widetilde{\mathcal{W}} = \mathcal{W}-\mathcal{J}
= (W-J) \otimes I$ the matrix
that describes how far is
$\mathcal{W}$ from the ideal matrix~$\mathcal{J}$.
 It can be shown that $\|\widetilde{\mathcal{W}}\| =
 \max\{\lambda_2,-\lambda_N\}=:\sigma \in [0,1)$. Next,
let $x^\bullet = \mathbf{1} \otimes x^\star$,
where we recall that $\mathbf{1}$ is
an all-ones vector (here of size $N \times 1$), and $x^\star$ is the ($d \times 1$) solution to \eqref{eqn-opt-prob-original}.
In other words, $x^\bullet$ concatenates
$N$ repetitions of $x^\star$ on top of each other;
 the $i$-th repetition
corresponds to node $i$ in the network.
Our goal is that, with a
distributed method, we have
$x^{(k)} \rightarrow x^{\bullet}$, or,
equivalently, $x_i^{(k)} \rightarrow x^{\star}$, for all
nodes $i=1,...,N$. Further, we define
 function $F:\,{\mathbb R}^{N d} \rightarrow \mathbb R$,
 by
 \begin{equation}
 F(x) := F(x_1,...,x_N)=\sum_{i=1}^N f_i(x_i).
 \end{equation}
Note that, under Assumption~1, $F$
is strongly convex with strong convexity parameter~$\mu$, and it
 has Lipschitz continuous gradient with Lipschitz constant~$L$.

\subsection{Review of distributed first order methods}
We review three existing distributed first order methods
that are of relevance to our studies --
the standard distributed
gradient method in~\cite{nedic_T-AC},
 the Extra method in~\cite{WotaoYinExtra},
 and the method in~\cite{Harnessing}.

\textbf{Standard distributed gradient method in~\cite{nedic_T-AC}}. The method updates
the solution estimate~$x_i^{(k)}$ at each node~$i$ by weight-averaging
 node $i$'s solution estimate with the estimates of its
immediate neighbors $j \in \Omega_i \setminus \{i\}$,
and then by taking a step in the negative local gradient's direction.
This corresponds to the following update rule at arbitrary node~$i$:
\begin{equation}
\label{eqn-starndard-DGM-node-i}
x_i^{(k+1)} = \sum_{j \in \Omega_i} W_{ij}\,x_j^{(k)} - \alpha\,\nabla f_i(x_i^{(k)}),\,\,k=0,1,...
\end{equation}
  In matrix format, the
network-wide update is as follows:\footnote{
To save space, we will
present all subsequent algorithms
in compact forms only. From a compact form,
it is straightforward to recover
local update forms at any node~$i$.
}
\begin{equation}
\label{eqn-starndard-DGM}
x^{(k+1)} = \mathcal{W}\,x^{(k)} - \alpha\,\nabla F(x^{(k)}),\,\,k=0,1,...
\end{equation}
%
 Here, constant $\alpha>0$ is the algorithm's step size.
A drawback of algorithm~\eqref{eqn-starndard-DGM}
is that (when constant step size $\alpha$ is used) it does not converge to
the exact solution $x^{\bullet}: =  \mathbf{1} \otimes x^\star$,
but only to a point in a $O(\alpha)$ solution neighborhood; see, e.g.,~\cite{WotaoYinDisGrad}.
 The algorithm can be made
 convergent to $x^\bullet$ by taking
 an appropriately set diminishing step size
 (e.g., square summable, non-summable), but the
  convergence rate of the resulting method is sublinear.

\textbf{Extra~\cite{WotaoYinExtra}}. The method modifies
the update rule~\eqref{eqn-opt-prob-original}
and achieves exact (global linear) convergence with a
fixed step size $\alpha$. Extra works as follows. It uses the following update rule:\footnote{
The Extra method utilizes two different weight matrices
in general. When these matrices are tuned as it is suggested in \cite{WotaoYinExtra},
Extra can be represented as in~\eqref{eqn-extra}--\eqref{eqn-extra-222}.
 The formulas here and in \cite{WotaoYinExtra} appear slightly different
 due to a different notation:
$\widetilde{W}=\frac{1}{2}(I+W)$ in \cite{WotaoYinExtra}
corresponds to $\mathcal{W}$ here.
It is required in \cite{WotaoYinExtra} for the analysis of \eqref{eqn-extra}--\eqref{eqn-extra-222}
that (in our notation) $(2\mathcal{W}-I)$
be doubly stochastic and that
 $\mathcal{W}$ be, besides being doubly stochastic, also positive definite.
The latter assumptions are not imposed here when analyzing the proposed method~\eqref{eqn-proposed}--\eqref{eqn-proposed-2} (Section~{IV}).}
{\allowdisplaybreaks{
\begin{eqnarray}
\label{eqn-extra}
x^{(1)} &=& \hspace{-2mm}\mathcal{W}\,x^{(0)} - \alpha \,\nabla F(x^{(0)})\\
\label{eqn-extra-222}
x^{(k+1)} &=&  \hspace{-2mm}
2\, \mathcal{W}\, x^{(k)} -
\alpha\,\nabla F(x^{(k)})- \mathcal{W}\,x^{(k-1)}  \\
&\,&+\,
\alpha\,\nabla F(x^{(k-1)}),\,\,\,k=1,2,... \nonumber
\end{eqnarray}}}
%

Reference~\cite{MokhtariDoubly}
demonstrates that algorithm~\eqref{eqn-extra}--\eqref{eqn-extra-222}
is a primal-dual gradient-like method; see, e.g., \cite{SaddlePointNedicAsu,SaddlePoint2,SaddlePointUzawa},
for primal-dual (sub)gradient methods. Denote by $\mathcal{L}:=I-\mathcal{W}$,
and consider the following constrained problem:
\begin{equation}
\label{eqn-equiv}
\mathrm{minimize}\,\,F(x)\,\,\,\,\,\,\mathrm{subject\,to}\,\,\,\,\,\,\frac{1}{\alpha}\mathcal{L}^{1/2}\,x=0.
\end{equation}
It can be shown, e.g.,~\cite{MokhtariDoubly},
that $\mathcal{L}^{1/2}\,x=0$, if and only
if $x_1=x_2=...=x_N$, where $x_i \in {\mathbb R}^d$
 is the $i$-th consecutive
 block of $x=(\,(x_1)^\top,...,(x_N)^\top\,)^\top$.
 Therefore,
 \eqref{eqn-equiv} is equivalent to
 \eqref{eqn-opt-prob-original}.
 Introduce the augmented Lagrangian function
  $\mathcal{A}:\,{\mathbb R}^{N d} \times {\mathbb R}^{N d} \rightarrow {\mathbb R}$
   (with the penalty parameter equal to $\alpha$)
   associated with \eqref{eqn-equiv}:
 \begin{equation}
 \label{eqn-AL}
 \mathcal{A}(x,\mu) = F(x) + \frac{1}{\alpha}\mu^\top \mathcal{L}^{1/2} x + \frac{1}{2 \alpha} x^\top \mathcal{L} x,
 \end{equation}
 and the corresponding dual problem:
 \begin{equation}
 \label{eqn-dual-prob}
 \mathrm{maximize}_{\mu \in {\mathbb R}^{N d}}\,\inf_{x \in {\mathbb R}^{Nd}} \mathcal{A}(x,\mu).
 \end{equation}
 %
 %
 Consider the following primal-dual method to solve~\eqref{eqn-dual-prob}:\footnote{More precisely,
 under appropriate conditions, with~\eqref{eqn-saddle-point-abstract} one has
 that $\left( x^{(k)},\,\mu^{(k)}\right)$ converges to
 a saddle point of function~$\mathcal{A}(x,\mu)$, which
 then implies that $x^{(k)}$ solves~\eqref{eqn-equiv} and~$\mu^{(k)}$ solves~\eqref{eqn-dual-prob}.}
 \begin{eqnarray}
 \label{eqn-saddle-point-abstract}
 x^{(k+1)} &=& x^{(k)} - \alpha\, \nabla_x \mathcal{A}(x^{(k)},\mu^{(k)})\\
 \mu^{(k+1)} &=&
 \mu^{(k)} +  \alpha\,\nabla_{\mu} \mathcal{A}(x^{(k+1)},\mu^{(k)}),k=0,1,...
 \end{eqnarray}
 with step size $\alpha>0$,
 arbitrary $x^{(0)} \in {\mathbb R}^{Nd}$, and $\mu^{(0)}=0$. Here, $\nabla_x$ and $\nabla_{\mu}$ denote the partial
 derivatives with respect to $x$ and $\mu$, respectively.
 Evaluating the partial derivatives,
 we arrive at the following method:
{\allowdisplaybreaks{
\begin{eqnarray*}
x^{(k+1)} &=&
x^{(k)} - \alpha \, \left( \frac{1}{\alpha}\mathcal{L}\,x^{(k)}  \right.
\\&\,& \left.\:\:\:\:\:+\,\, \nabla F(x^{(k)})
+\frac{\mathcal{L}^{1/2}}{\alpha}\mu^{(k)} \right) \\
\mu^{(k+1)}
&=&
\mu^{(k)} + \mathcal{L}^{1/2}\,x^{(k+1)},k=0,1,...
\end{eqnarray*}}}
Introducing the new
variable $u^{(k)}:=\frac{1}{\alpha}\mathcal{L}^{1/2}\,\mu^{(k)}$, one arrives
at the following method:\footnote{See also~\cite{LinearRateAL} for similar methods with multiple primal updates per each dual update.}
\begin{eqnarray}
\label{eqn-saddle-extra}
x^{(k+1)} &=&
x^{(k)} - \alpha \, \left( \frac{1}{\alpha}\mathcal{L}\,x^{(k)} + \nabla F(x^{(k)}) +  u^{(k)} \hspace{-1mm}\right) \\
\label{eqn-saddle-extra-2}
u^{(k+1)}
&=&
u^{(k)} + \frac{1}{\alpha}\mathcal{L}\,x^{(k+1)},k=0,1,...
\end{eqnarray}
It turns out that \eqref{eqn-saddle-extra}--\eqref{eqn-saddle-extra-2} with a proper initialization is equivalent to \eqref{eqn-extra}--\eqref{eqn-extra-222}.\footnote{Reference~\cite{MokhtariDoubly}
shows the equivalence under a more general initialization; we keep the one as in Lemma \ref{lemma-DSA-Mokhtari-proof} for simplicity.}
%
%
\begin{lemma}[\cite{MokhtariDoubly}]
\label{lemma-DSA-Mokhtari-proof}
The sequence of iterates $\{x^{(k)}\}$ generated by algorithm~\eqref{eqn-extra}--\eqref{eqn-extra-222}, with initialization $x_i^{(0)}=x_j^{(0)}$, for all $i,j$,
is the same as the sequence of iterates generated by \eqref{eqn-saddle-extra}--\eqref{eqn-saddle-extra-2},
with the same initialization of $x_i^{(0)}$, $i=1,...,N$, and
with $u^{(k)}$ initialized to zero.
\end{lemma}
%
%
 Under Assumptions 1 and 3 an appropriately chosen step size
 $\alpha$,
 one has that $x^{(k)} \rightarrow x^{\bullet}$, and
 $u^{(k)} \rightarrow -\nabla F(x^{\bullet})$,
 at an R-linear rate~\cite{WotaoYinExtra}.

\textbf{The exact method in~\cite{Harnessing}}. The authors of~\cite{Harnessing}, see also~\cite{SmallGainNedicUncoord,SmallGainNedicTimeVar,SmallGainKin1,SmallGainKin2,HarnessingNesterov}, consider a distributed
first order method that, besides solution estimate $x^{(k)} \in {\mathbb R}^{Nd}$,
also maintains an auxiliary variable~$s^{(k)} \in {\mathbb R}^{Nd}$.
 Here, at each node $i$, quantity $s_i^{(k)} \in {\mathbb R}^d$
 serves to approximate the network-wide gradient average $\frac{1}{N}\sum_{i=1}^N
 \nabla f_i(x_i^{(k)})$; then, the gradient
 contribution
 $-\alpha \nabla F(x^{(k)})$
  with the standard distributed gradient method \eqref{eqn-starndard-DGM}
  is replaced with $s^{(k)}$. More precisely, the update rule for
  $k=0,1,...$ is as follows:
\begin{eqnarray}
\label{eqn-harnessing}
x^{(k+1)} &=& \mathcal{W}\,x^{(k)} - \alpha\,s^{(k)}\\
\label{eqn-harnessing-primeprime}
s^{(k+1)} &=& \mathcal{W}\,s^{(k)} + \nabla F(x^{(k+1)}) - \nabla F(x^{(k)}),
\end{eqnarray}
with $s^{(0)} = \nabla F(x^{(0)})$.
  The method also achieves
 a global R-linear convergence
 under appropriately
 chosen step-size
 $\alpha$, when
 Assumptions~1 and~3 are in force.
\section{The proposed method}
Subsection~{III-A} presents the method that we propose and
explains how to recover the existing methods in~\cite{WotaoYinExtra,Harnessing}
 by a particular setting of a proposed method's parameter matrix. Subsection~{III-B}
gives further insights into the proposed method and explains how to
tune the parameter matrix. Finally, Subsection~{III-C} provides
a primal-dual interpretation of the proposed method and the methods in~\cite{WotaoYinExtra,Harnessing}.

%
\subsection{The proposed method  and its relation with existing algorithms}
We now describe the generalized exact first order method that we propose.
The method subsumes the known methods~\cite{WotaoYinExtra}
and~\cite{Harnessing} upon a
specific choice of the tuning parameters, as explained below.
The algorithm maintains over iterations $k$
 the primal variable (solution estimate)
 $x^{(k)} \in {\mathbb R}^{N d}$  and
  the dual variable $u^{(k)} \in {\mathbb R}^{Nd}$, initialized
  with the zero vector. The update rule is for $k=0,1,...$ given as follows:
\begin{eqnarray}
\label{eqn-proposed}
x^{(k+1)} &=&
\mathcal{W}\,x^{(k)} - \alpha \, \left( \nabla F(x^{(k)}) + u^{(k)} \right) \\
\label{eqn-proposed-2}
u^{(k+1)}
&=&
u^{(k)} - \mathcal{L}\, \left( \nabla F(x^{(k)}) + u^{(k)} - \mathcal{B}\, x^{(k)}\right).
\end{eqnarray}
Here, quantities $\mathcal{W}$, $\mathcal{L}$, and $\alpha$
 are the same as before.
 We note that one can use two
 different (doubly stochastic) weight matrices
 $\mathcal{W}_1$ and $\mathcal{W}_2$ in
 \eqref{eqn-proposed} and
 \eqref{eqn-proposed-2}
 above; more precisely,
 one can replace
 $\mathcal{W}$ with $\mathcal{W}_1$ in
 \eqref{eqn-proposed} and
 $\mathcal{L}$ with $I-\mathcal{W}_2$ in
 \eqref{eqn-proposed-2}. However,
 throughout we keep a single weight matrix~$\mathcal{W}$
  for simplicity of the analysis and presentation.
 Quantity
   $\mathcal{B}$ is a $(Nd) \times (Nd)$
     symmetric matrix that respects the block-sparsity
    pattern of the underlying graph $\mathcal{G}$ and satisfies the property that,
    for any $y \in {\mathbb R}^d$, there exists some $c \in {\mathbb R}$,
     such that $\mathcal{B}\,(\mathbf{1}\otimes y) = c\,(\mathbf{1}\otimes y)$.
     Specifically, we will consider the following choices (that clearly obey the latter conditions): 1) $\mathcal{B} = b\,I$,
    where $b \geq 0$ is a scalar parameter;
    and 2) $\mathcal{B} = b^\prime\,\mathcal{W}$,
    for $b^\prime \geq 0$.
    These choices are easy to implement and  incur no additional
    communication overhead while lead to efficient algorithms (see also Section~{V}).
    That is, with the above choices of matrix $\mathcal{B}$, the proposed
    method \eqref{eqn-proposed}--\eqref{eqn-proposed-2} requires communicating
    $2$ vectors of size $d$ per node, per iteration.
    We note that, for a generic choice of matrix $\mathcal{B}$,
     method \eqref{eqn-proposed}--\eqref{eqn-proposed-2}
     incurs one additional communication (of a $d$-dimensional vector) per node,
     per iteration, i.e., it requires communicating
     3 vectors of size $d$ per node, per iteration.


The next lemma, proved in the Appendix, explains how to recover the existing
exact distributed first order methods
from~\eqref{eqn-proposed}--\eqref{eqn-proposed-2}.\footnote{The equivalence claimed in Lemma~\ref{lemma-equivalence}
is in the sense that the methods generate the same
sequence of iterates $\{x^{(k)}\}$ under the same initialization $x^{(0)}$, the
appropriate initialization of methods' auxiliary variables, and the
same step size~$\alpha$.}
\begin{lemma} Consider algorithm~\eqref{eqn-proposed}--\eqref{eqn-proposed-2}.
\label{lemma-equivalence}
 Then, the following  holds.
 \begin{itemize}
 \item [(a)]
 Algorithm~\eqref{eqn-proposed}--\eqref{eqn-proposed-2} with  $\mathcal{B} = 0$ is equivalent to method~\eqref{eqn-harnessing}, proposed in~\cite{Harnessing}.
 \item [(b)]
 Algorithm~\eqref{eqn-proposed}--\eqref{eqn-proposed-2} with
 $\mathcal{B} = \frac{1}{\alpha} \mathcal{W}$ is equivalent to method~\eqref{eqn-extra}--\eqref{eqn-extra-222}, proposed in~\cite{WotaoYinExtra}.
 \end{itemize}
 \end{lemma}
%
%
%
%

\subsection{Further insights into the proposed method and parameter tuning}
We give further intuition and insights into the
proposed method, the existing algorithms in~\cite{Harnessing}
 and~\cite{WotaoYinExtra},
 and we also describe how
 to improve upon existing
 methods by appropriately setting
 matrix~$\mathcal{B}$.
 Denote by
 $e_x^{(k)}:=x^{(k)} - x^{\bullet}$
  and $e_u^{(k)}:=u^{(k)}+\nabla F(x^{\bullet})$
   the primal and dual errors, respectively. Also,
   let the primal-dual error vector
 $e^{(k)}:=\left( (e_x^{(k)})^\top, (e_u^{(k)})^\top\right)^\top$, and
$\mathcal{H}_k :=\int_{t=0}^1 \nabla^2 F\left(x^\bullet + t \,(x^{(k)}-x^\bullet)\right)\,dt$.
 We have the following Lemma.\footnote{In Section~{IV}, we show
 that, under Assumptions~1 and~3, $e^{(k)}$ converges to zero R-linearly.}
 \begin{lemma}
 \label{lemma-recursion}
 Let Assumptions~1--3 hold, and
 consider algorithm~\eqref{eqn-proposed}--\eqref{eqn-proposed-2}.
 Then, the primal-dual error vector
 $e^{(k)}:=\left( (e_x^{(k)})^\top, (e_u^{(k)})^\top\right)^\top$,
 for $k=0,1,...,$
 satisfies the following recursion:
\begin{equation}
\begin{bmatrix}
    e_x^{(k+1)}   \\
    e_u^{(k+1)}
\end{bmatrix} =
\begin{bmatrix}
    \mathcal{W}-\alpha\,\mathcal{H}_k & -\alpha\,I \\
    \left( \mathcal{W}-I\right)\left( \mathcal{H}_k-\mathcal{B}\right)
    & \mathcal{W}-\mathcal{J}
\end{bmatrix}
\,
\begin{bmatrix}
    e_x^{(k)}   \\
    e_u^{(k)}
\end{bmatrix}.
\end{equation}
 \end{lemma}
Lemma \ref{lemma-recursion}
expresses the primal dual error $e^{(k)}$
 as a recursion, guided by a time
 varying matrix:
 \begin{equation}
 \label{eqn-error-matrix}
 \mathcal{M}_k:=\begin{bmatrix}
    \mathcal{W}-\alpha\,\mathcal{H}_k & -\alpha\,I \\
    \left( \mathcal{W}-I\right)\left( \mathcal{H}_k-\mathcal{B}\right)
    & \mathcal{W}-\mathcal{J}
\end{bmatrix},
 \end{equation}
that we refer to as the error dynamics matrix.\footnote{In Section~{IV},
we show that, under appropriate choice of
parameters $\alpha$ and $\mathcal{B}$,
the primal-dual error $e^{(k)}$ converges
to zero R-linearly.}
 Equation~\eqref{eqn-error-matrix} shows the partitioning of the error dynamics matrix as a $2 \times 2$ block matrix
with blocks of size $(Nd) \times (Nd)$.
 The $(1,1)$-th block $(\mathcal{W}-\alpha\,\mathcal{H}_k)$
 describes
 how the current primal error
  affects the next primal error,
  the $(1,2)$-th block $(-\alpha\,I)$
   describes
 how the current dual error
  affects the next primal error, and so on.
    Specifically,
    with the methods in~\cite{Harnessing}
    and~\cite{WotaoYinExtra},
 the blocks on the positions $(1,1)$, $(1,2)$, and $(2,2)$
   are of the same structure for both
  methods, and are of the same structure as in~\eqref{eqn-error-matrix}.
  For clarity of explanation,
  assume that
  the $f_i$'s are strongly convex quadratic functions,
  so that $\mathcal{H}_k = \mathcal{H} = \nabla^2 F(x) = \mathrm{const}$,
  for any $x \in {\mathbb R}^{Nd}$,
  with $\mu\,I \preceq \mathcal{H} \preceq L\,I$, and
  of course the same $\mathcal{H}$ appearing in the error dynamics matrices for each of the three methods.
    Then, the blocks $(1,1)$, $(1,2)$, and $(2,2)$ match completely
  for the three methods.
  However, the error dynamics matrices for different methods differ in the $(2,1)$-th block.
  Specifically, with~\cite{Harnessing},
  the $(2,1)$-th block of the corresponding error dynamics matrix equals
  $\left( \mathcal{W} - I\right)\,\mathcal{H}$.
  On the other hand, with \cite{WotaoYinExtra}
  the block equals
   $\left( \mathcal{W} - I\right)\left( \mathcal{H} - \frac{1}{\alpha}\mathcal{W}\right).$
    Intuitively,
    one may expect that if
    $[\mathcal{M}_k]_{2,1}$ is
    smaller (as measured by an appropriate matrix norm),
     then the algorithm's convergence is likely to be faster.\footnote{This
     intuition is corroborated more formally in Theorem~\ref{theorem-R-linear-rate} and Remark~4.}
    This provides an intuition
    on the comparison between
    \cite{Harnessing}
     and~\cite{WotaoYinExtra}.
      Namely, if $\mathcal{H}$
      is small relative to
      $(\mathcal{H} -\frac{1}{\alpha}\,\mathcal{W})$ (for instance,
      when $\alpha$ is very small and so $\frac{1}{\alpha}\,\mathcal{W}$ has a very large norm),
       then we expect
       that the method in~\cite{Harnessing}
         is faster than the method in~\cite{WotaoYinExtra},
         and vice versa.
          This intuition is
         confirmed in Section~{V}
          by numerical examples.

We can go one step further and
 seek to tune matrix $\mathcal{B}$
  such that $[\mathcal{M}_k]_{2,1} =
  (\mathcal{W}-I)(\mathcal{H}-\mathcal{B})$
   is smallest in an appropriate sense.
   We consider separately
   the cases $\mathcal{B}=b\,I$ and
   $\mathcal{B}=b^\prime \,\mathcal{W}$.
   For the former, a possible worst case-type approach is as
 follows:
 choose parameter $b$
 that solves the following problem:
 \begin{equation}
 \label{eqn-21-block}
 \mathrm{min}_{\,b \geq 0}\,\left\{\sup_{\mathcal{H} \in \mathbb{H}}\|\mathcal{H}-b\,I\|\right\},
 \end{equation}
 where $\mathbb{H}$ is the
 set of all $(Nd)\times (Nd)$ symmetric
 block diagonal matrices $\mathcal{H}$ with
 arbitrary $d \times d$ diagonal blocks
 that in addition obey the following condition:
 $\mu\,I \preceq \mathcal{H} \preceq L\,I$.
 %
%
 It is easy to show (see the Appendix for details)
 that the solution to \eqref{eqn-21-block} is $b^\star = \frac{\mu+L}{2}$.
 Note that the choice
 $\mathcal{B}=\frac{\mu+L}{2}\,I$
 in~\eqref{eqn-proposed}--\eqref{eqn-proposed-2}
  does not match either \cite{Harnessing}
  or \cite{WotaoYinExtra}
  method and thus represents
  a novel algorithm.
  For the latter choice
  $\mathcal{B}=b^\prime\,\mathcal{W}$,
  the analogous problem:
 \begin{equation}
 \label{eqn-21-block-cal-W}
 \mathrm{min}_{\,b^{\prime} \geq 0}\,\left\{\sup_{\mathcal{H} \in \mathbb{H}}\|\mathcal{H}-b^{\prime}\,\mathcal{W}\|\right\}
 \end{equation}
 is more challenging.
 A sub-optimal choice for $\lambda_N >0$, as shown
 in the Appendix, is
 $b^{\prime}=\frac{L+\mu}{1+\lambda_N}$,
 where we recall
 that $\lambda_N$
  is the smallest eigenvalue of matrix~$W$.
 Extensive simulations (see Also Section~V) show
 that the simple choice $b^{\prime}=L$ works well in practice.

Note that, ideally, one would like to
   minimize with respect to $b$ the following quantity:
   $\sup_{\mathcal{H}:\,\mu\,I \preceq \mathcal{H} \preceq L\,I}\|\mathcal{M}_k\|$,
   i.e., one wants to take into account the full matrix $\mathcal{M}_k$.
   This problem is challenging in general and is hence
   replaced here by method~\eqref{eqn-21-block} (or, similarly, (21)), i.e., by considering the $(2,1)$-th block of
   $\mathcal{M}_k$ only. Note that a smaller norm of the $(2,1)$-th block of
   $\mathcal{M}_k$ might not necessarily imply a smaller norm of the full matrix
   $\mathcal{M}_k$. However, extensive
   numerical experiments on quadratic and logistic losses
   (see also Section~5)  demonstrate
   that tuning method~\eqref{eqn-21-block} yields
   fast algorithms while at the same time is very cheap.

\subsection{Primal-dual interpretations}
We now provide a primal-dual interpretation
of the proposed method.
 The interpretation
 builds on construction~(8)--(11) from~\cite{MokhtariDoubly}.
 It is worth noting
 that~\cite{SmallGainNedicTimeVar}
 provides a primal-dual interpretation
 of the method therein, equivalent to the method in~\cite{Harnessing}.
  The construction in~\cite{SmallGainNedicTimeVar}
  starts from a different problem reformulation than~(8)
   and utilizes a different quadratic penalty term for
   the Lagrangian function.

To start, we write
the methods \eqref{eqn-extra}--\eqref{eqn-extra-222} and \eqref{eqn-harnessing}--\eqref{eqn-harnessing-primeprime}
in another equivalent form (see the Appendix as to why this
equivalence also holds.) Namely,
\eqref{eqn-extra}--\eqref{eqn-extra-222}
can be equivalently represented as follows
(this is essentially a re-write of \eqref{eqn-saddle-extra}--\eqref{eqn-saddle-extra-2})
but is useful to present it here):
{\allowdisplaybreaks{
\begin{eqnarray}
\label{eqn-saddle-extra-new}
x^{(k+1)} &=&
\mathcal{W}\,x^{(k)} - \alpha \, \left( \nabla F (x^{(k)}) + u^{(k)} \right) \\
\label{eqn-saddle-extra-new-2}
u^{(k+1)}
&=&
u^{(k)} + \frac{1}{\alpha}\mathcal{L}\,x^{(k+1)},k=0,1,...
\end{eqnarray}}}
Similarly, \eqref{eqn-harnessing}
can be, for $k=0,1,...$, equivalently represented as:
\begin{eqnarray}
\label{eqn-harness-new}
x^{(k+1)} &=&
\mathcal{W}\,x^{(k)} - \alpha \, \left( \nabla F (x^{(k)}) + u^{(k)} \right) \\
\label{eqn-harness-new-2}
u^{(k+1)}
&=&
u^{(k)} + \frac{1}{\alpha}\mathcal{L}\, x^{(k+1)} - \frac{1}{\alpha}\mathcal{W}\,\mathcal{L}\,x^{(k)}.
\end{eqnarray}
We can re-interpret \eqref{eqn-harness-new}
as a primal-dual gradient-like method for solving~\eqref{eqn-dual-prob}.
Namely, due to the fact that matrices
$\mathcal{W}$ and $\mathcal{L}$ commute, it is easy to see that \eqref{eqn-harness-new}--\eqref{eqn-harness-new-2}
corresponds to the following method:
{\allowdisplaybreaks{
 \begin{eqnarray}
 \label{eqn-saddle-point-abstract-2}
 x^{(k+1)} &=& x^{(k)} - \alpha\, \nabla_x \mathcal{A}(x^{(k)},\mu^{(k)})\\
 \mu^{(k+1)} &=&
 \mu^{(k)} +  \alpha\,\nabla_{\mu} \mathcal{A}(x^{(k+1)},\mu^{(k)})\\
 &\,&-
 \alpha\,\mathcal{W}\,\nabla_{\mu} \mathcal{A}(x^{(k)},\mu^{(k)}). \nonumber
 \end{eqnarray}}}
Hence,~\eqref{eqn-harness-new}
 is a primal-dual gradient-like method that modifies
 the dual update step to also
 incorporate the (weighted) previous dual gradient term.
 Clearly, the
 proposed generalized method~\eqref{eqn-proposed}--\eqref{eqn-proposed-2}
 also
 incorporates the (weighted) previous dual gradient term.
 It is shown in the Appendix that, assuming
 that matrices $\mathcal{B}$ and $\mathcal{L}$ commute (which
 is the case for the two specific choices
 $\mathcal{B}=b\,I$ and $\mathcal{B}=b^\prime\,\mathcal{W}$ considered here),
 we have that~\eqref{eqn-proposed}--\eqref{eqn-proposed-2} is equivalent to the following primal-dual method:
  {\allowdisplaybreaks{
 \begin{eqnarray}
 \label{eqn-saddle-point-abstract-prop}
 x^{(k+1)} &=& x^{(k)} - \alpha\, \nabla_x \mathcal{A}(x^{(k)},\mu^{(k)})\\
  \label{eqn-saddle-point-abstract-prop-2}
 \mu^{(k+1)} &=&
 \mu^{(k)} +  \alpha\,\nabla_{\mu} \mathcal{A}(x^{(k+1)},\mu^{(k)})\\
 &\,&-
 \alpha\,\left(\mathcal{W}-\alpha\,\mathcal{B}\right)\,\nabla_{\mu} \mathcal{A}(x^{(k)},\mu^{(k)}). \nonumber
 \end{eqnarray}}}

\section{Convergence rate analysis}
This Section presents the results on
global R-linear convergence of the proposed
method~\eqref{eqn-proposed}--\eqref{eqn-proposed-2}, and provides
the needed intermediate results and proofs.
The Section is organized as follows.
First, Subsection~{IV-A} states
the result (Theorem~\ref{theorem-R-linear-rate}) on global R-linear convergence
of the proposed method~\eqref{eqn-proposed}--\eqref{eqn-proposed-2}
and discusses the implications of the result.
Subsection~{IV-B} sets up the analysis
and gives preliminary Lemmas.
Finally, Subsection~{IV-C}
proves a series of intermediate Lemmas,
followed by the proof of Theorem~\ref{theorem-R-linear-rate}.

\subsection{Global R-linear convergence rate: Statement of the result}
We show that, under appropriate choice
of step size $\alpha$,
the proposed method~\eqref{eqn-proposed}--\eqref{eqn-proposed-2}
converges to the exact solution at a global R-linear rate.
Recall quantity $\sigma = \max\{\lambda_2,-\lambda_N\} \in [0,1)$,
with $\lambda_i$ the $i$-th largest eigenvalue of~$W$.

 \begin{theorem}
 \label{theorem-R-linear-rate}
 Consider algorithm~\eqref{eqn-proposed}--\eqref{eqn-proposed-2} with $\mathcal{B}=b\,I$,
 and let Assumptions~1 and~3 hold.
 Further, let
  $\alpha < \min
  \left\{\frac{(1-{\sigma})\,\mu}{19\,L^2},\,\frac{(1-{\sigma})^2\,\mu}{192\,L^\prime\,L}\right\}$,
  where $L^\prime=\left( L^2+b^2-2\,b\,\mu\right)^{1/2}$.
 Then, the sequence
 of iterates
 $x^{(k)}$
 generated by algorithm~\eqref{eqn-proposed}--\eqref{eqn-proposed-2}
 converges to $x^\bullet  = \mathbf{1} \otimes x^\star$
 R-linearly, i.e.,
 there holds
 $\|x^{(k)}-x^\bullet\|=O\left( r^k\right)$,
 with $r \in (0,1)$.
  The
 convergence factor~$r$
  is at most $\left(\max\{1-\frac{\alpha\,\mu}{2},\,\frac{1+{\sigma}}{2}\}+\epsilon\right)$,
   where $\epsilon>0$ is arbitrarily small.
 \end{theorem}

Several Remarks on Theorem \ref{theorem-R-linear-rate} are now in order.

\begin{remark}
When $\mathcal{B}=b\,I$ is replaced
with a generic symmetric matrix $\mathcal{B}$
that respects the sparsity pattern of graph~$\mathcal{G}$
 and satisfies the property that,
    for any $y \in {\mathbb R}^d$, there exists some $c \in {\mathbb R}$,
     such that $\mathcal{B}\,(\mathbf{1}\otimes y) = c\,(\mathbf{1}\otimes y)$, Theorem~\ref{theorem-R-linear-rate}
  continues to hold
  with $L^\prime$ replaced with
  constant $\left( L + \|\mathcal{B}\| \right)$;
  see the Appendix for the proof.
\end{remark}

\begin{remark}
The maximal admissible step size
for which
Theorem \ref{theorem-R-linear-rate}
guarantees R-linear
convergence
 can be very small for
 poorly conditioned problems ($L/\mu$ large) and/or
 weekly connected networks (${\sigma}$ close to one).
 However,
extensive simulations
show that the proposed method
converges (R-linearly) in practice with large step sizes, e.g.,
$\alpha=\frac{1}{3\,L}$.
 Similar theoretical and practical
 admissible values for step size
 $\alpha$ are reported in earlier
 works for the methods studied therein~\cite{Harnessing,WotaoYinExtra}.
\end{remark}

\begin{remark} Theorem~\ref{theorem-R-linear-rate}
generalizes existing R-linear convergence rate
results of exact distributed first order methods, e.g., \cite{Harnessing},
to a wider class of algorithms.
Recall that for $b=0$
we recover the method in~\cite{Harnessing}. Note that
for $b=0$ we have $L^{\prime}=L$.
 Abstracting universal constants, the convergence factor obtained in~\cite{Harnessing}
  (see Theorem~1 and Lemma~2 therein)
 and the one obtained here
 are the same and equal
 \begin{equation}
 \label{eqn-bound-conv-factor-new}
 1-\Omega\left( \frac{(1-{\sigma})^2\mu^2}{L^2}\right).
 \end{equation}
 This bound is obtained here by setting the
 maximal step size $\alpha $ permitted by Theorem~\ref{theorem-R-linear-rate},
 which is $\alpha=\Omega\left( \frac{(1-\sigma)^2\,\mu}{L^2}\right)$.
\end{remark}

\begin{remark} Adapting the results in~\cite{SmallGainNedicUncoord} to our
 setting by letting the nodes' step-sizes and their strong convexity parameters be equal,
 the results from~\cite{SmallGainNedicUncoord} yield the convergence factor:
 \[
 1 - \Omega\left( \frac{(1-\sigma)^2}{\sqrt{N}}\frac{\mu^{3/2}}{L^{3/2}} \right).
 \]
Comparing this with bound \eqref{eqn-bound-conv-factor-new}
obtained here, we can see that our bound
 exhibits better
scaling with respect to the number of nodes~$N$;
on the other hand, the bounds in~\cite{SmallGainNedicUncoord,SmallGainNedicTimeVar}
 exhibit better scaling in terms of condition number~$(L/\mu)$.
\end{remark}

\begin{remark}
It is interesting to observe what happens in Theorem~\ref{theorem-R-linear-rate} when $b=\mu$ --
  the case that corresponds to a novel exact method.
 Then, one has $L^{\prime}=\left( L^2-\mu^2\right)^{1/2}$,
 which is arbitrarily small when $L$ becomes close to $\mu$.
 Hence, for well-conditioned problems
 ($L$ close to $\mu$),
 the maximal admissible step size
 in Theorem~\ref{theorem-R-linear-rate} becomes
 $\alpha=\Omega\left( \frac{(1-{\sigma})\mu}{L^2}\right)$,
 and the corresponding
 convergence factor is
 $1-\Omega\left( \frac{(1-{\sigma})\,\mu^2}{L^2}\right)$.
 Hence, according to
 the upper bounds derived here and in \cite{Harnessing},
 the proposed method with $b=\mu$
 improves convergence factor over \cite{Harnessing}
 from $1-\Omega\left( \frac{(1-{\sigma})^2\mu^2}{L^2}\right)$ to $1-\Omega\left( \frac{(1-{\sigma})\,\mu^2}{L^2}\right)$ for
 well-conditioned problems ($L$ sufficiently close to $\mu$).
 Simulations confirm large gains in
 convergence speed of the new method over~\cite{Harnessing} (see Section~{V}).
\end{remark}

\subsection{Setting up analysis and preliminary Lemmas}
The proof of Theorem~\ref{theorem-R-linear-rate}
is based on the small gain Theorem~\cite{SmallGainBook}.
  It is worth noting that analytical approaches
using small gain theorem have been developed in~\cite{SmallGainNedicUncoord} and \cite{SmallGainNedicTimeVar} prior to our work and
hold for a more general network and step-size models than what we study here. Namely,
reference~\cite{SmallGainNedicTimeVar} carries out analysis under undirected and directed time varying graphs,
while reference~\cite{SmallGainNedicUncoord} allows for node-dependent step-sizes.
However, the analysis here is different. First, it is carried out
for a different algorithm in general (the proposed method \eqref{eqn-proposed}--\eqref{eqn-proposed-2} that involves matrix $\mathcal{B}$);
as such, it reveals the novel insight that the
negative coupling from the primal to the dual error
can be partially compensated by tuning matrix~$\mathcal{B}$;
see also Remark~5. Second,
when the results here are specialized to the method in~\cite{SmallGainNedicUncoord,SmallGainNedicTimeVar},
the obtained convergence factors are different. (See Remark~4 for details.)

Besides the small gain Theorem, this  Subsection also reviews some known results on the
  convergence of inexact (centralized)
  gradient methods that will be used subsequently.

We start by reviewing
the setting of the small gain theorem~\cite{SmallGainBook}.
 Denote by
 $\mathbf{a}:=a^{(0)},a^{(1)},...,a^{(k)},...$
  an infinite sequence of vectors,
  $a^{(k)} \in {\mathbb R}^p$, $k=0,1,...$
  For a fixed~$\delta \in (0,1)$,
  define the following quantities:
  \begin{eqnarray}
  \label{eqn-metrics}
  \|\mathbf{a}\|^{\delta,K} &:=& \max_{k=0,...,K}\left\{\frac{1}{\delta^k}\|a^{(k)}\|\right\}\\
  \|\mathbf{a}\|^{\delta} &:=& \sup_{k \geq 0}\left\{\frac{1}{\delta^k}\|a^{(k)}\|\right\}.
  \end{eqnarray}
Clearly,
we have that
$\|\mathbf{a}\|^{\delta,K}
 \leq \|\mathbf{a}\|^{\delta,K^\prime} \leq  \|\mathbf{a}\|^{\delta}$,
 for $K^\prime \geq K \geq 0$.
It is also clear that, if $\|\mathbf{a}\|^{\delta}$ is finite,
then the sequence $a^{(k)}$ converges
to zero R-linearly. Indeed,
provided that $\|\mathbf{a}\|^{\delta} \leq C_{a}<\infty$,
there holds:
 $
\|a^{(k)}\| \leq C_{a} \,\delta^k,\,\,k=0,1,...
 $
Hence, if for some $\delta \in (0,1)$ we have
that $\|\mathbf{a}\|^{\delta}$ is finite,
 then
the sequence $\{a^{(k)}\}$ converges
to zero R-linearly with
convergence factor at most $\delta$.
 We state the small gain theorem
 for two infinite sequences
 $\mathbf{a}$ and $\mathbf{b}$
 as this suffices for our analysis;
 for the more general Theorem
 involving an arbitrary (finite)
 number of infinite sequences, see, e.g.,~\cite{SmallGainBook,SmallGainNedicUncoord}.

 \begin{theorem}
 \label{theorem-small-gain}
 Consider two infinite sequences
 $\mathbf{a}=a^{(0)},a^{(1)},...$, and
 $\mathbf{b}=b^{(0)},b^{(1)},...$,
 with $a^{(k)}$, $b^{(k)}\in {\mathbb R}^p$,
 $k=0,1,...$
 Suppose that for some $\delta \in (0,1)$, and for all $K=0,1,...$, there holds:
 \begin{eqnarray}
 \label{eqn-thm-small-gain-1}
 \|\mathbf{a}\|^{\delta,K} &\leq& \gamma_1\, \|\mathbf{b}\|^{\delta,K} + \omega_1\\
 \label{eqn-thm-small-gain-2}
 \|\mathbf{b}\|^{\delta,K} &\leq& \gamma_2\, \|\mathbf{a}\|^{\delta,K} + \omega_2,
 \end{eqnarray}
 where
 $\gamma_1,\gamma_2 \geq 0$ and $\gamma_1\,\gamma_2 <1.$
 Then, there holds:
 \begin{equation}
 \label{eqn-small-gain-claim}
 \|\mathbf{a}\|^{\delta } \leq \frac{1}{1-\gamma_1\,\gamma_2} \left( \omega_2\,\gamma_1 + \omega_1\right).
 \end{equation}
 \end{theorem}

 We will frequently use the
 following simple Lemma.
\begin{lemma}
 \label{lemma-small-gain}
 Consider two infinite sequences
 $\mathbf{a}=a^{(0)},a^{(1)},...$, and
 $\mathbf{b}=b^{(0)},b^{(1)},...$,
 with $a^{(k)}$, $b^{(k)}\in {\mathbb R}^p$.
 Suppose that, for all $k=0,1,...,$ there holds:
 \begin{equation}
 \label{eqn-lemma-aux-small-gain}
 \|a^{(k+1)}\|
 \leq
 c_1\,\|a^{(k)}\|
 +
 c_2\,\|b^{(k)}\|,
 \end{equation}
 where $c_i \geq 0$, $i=1,2.$
  Then, for all $K =0,1,...$,
  for any $\delta \in (0,1)$, we have:
  \begin{equation}
 \label{eqn-lemma-aux-small-gain-pp}
 \|\mathbf{a}\|^{\delta,K}
 \leq
 \frac{c_1}{\delta} \,
 \|\mathbf{a}\|^{\delta,K}
 +
 \frac{c_2}{\delta}\,\|\mathbf{b}\|^{\delta,K} + \|a^{(0)}\|.
 \end{equation}
 \end{lemma}

\begin{IEEEproof}
Divide inequality \eqref{eqn-lemma-aux-small-gain} by
$\frac{1}{\delta^{k+1}}$, $\delta \in (0,1)$. The resulting inequality implies, for
all $K=1,2,...$ that:
{\small{
{\allowdisplaybreaks{
 \begin{eqnarray}
 &\,&\max_{k=0,...,K-1}\left\{\frac{1}{\delta^{k+1}}\|a^{(k+1)}\|\right\} 
 \leq
 \frac{c_1}{\delta}\,\max_{k=0,...,K-1}\left\{\frac{1}{\delta^k}\|a^{(k)}\|\right\} \nonumber \\
 &+&\,\,
 \frac{c_2}{\delta}\,\max_{k=0,...,K-1}\left\{\frac{1}{\delta^k}\|b^{(k)}\|\right\} 
 =
 \frac{c_1}{\delta} \|\mathbf{a}\|^{\delta,K-1}
 +\frac{c_2}{\delta} \|\mathbf{b}\|^{\delta,K-1}  \nonumber\\
 &\,&\,\,\leq
 \label{eqn-lemma-simple-proof-111}
 \frac{c_1}{\delta} \|\mathbf{a}\|^{\delta,K}
 +\frac{c_2}{\delta} \|\mathbf{b}\|^{\delta,K}.
 \end{eqnarray}}}}}
Note that~\eqref{eqn-lemma-simple-proof-111} implies that, for all $K=0,1,...$
{\allowdisplaybreaks{
\begin{eqnarray*}
&\,&\|\mathbf{a}\|^{\delta,K} =
\max_{k=-1,0,...,K-1}\left\{ \frac{1}{\delta^{k+1}}\|a^{(k+1)}\|\right\} \\
 &\,&\,\,\,\, \leq\,\,
\frac{c_1}{\delta} \|\mathbf{a}\|^{\delta,K}
 +\frac{c_2}{\delta} \|\mathbf{b}\|^{\delta,K}
  +\|a^{(0)}\|,
\end{eqnarray*}}}
which is precisely what we wanted to show.
\end{IEEEproof}
The use of the Lemma
will be to bound
$\|a\|^{\delta,K}$
 by $\|b\|^{\delta,K}$.
 Namely, whenever
 $c_3:=\frac{c_1}{\delta}<1$,
 \eqref{eqn-lemma-aux-small-gain} implies the following bound:
 \begin{equation}
 \label{eqn-arrow}
 \|\mathbf{a}\|^{\delta,K}
 \leq
 \frac{c_2/\delta}{1-c_3}\,\|\mathbf{b}\|^{\delta,K}
  + \frac{1}{1-c_3}\,\|a^{(0)}\|.
 \end{equation}
 We will also need the following
 Lemma from~\cite{Schmidt} on the convergence
 of inexact (centralized) gradient methods.
 \begin{lemma}
 \label{lemma-inexact}
 Consider unconstrained minimization
 of function $\phi:\,{\mathbb R}^p \rightarrow \mathbb R$,
 where $\phi$ is assumed to be
 strongly convex with strong
 convexity parameter $m$, and it also
 has Lipschitz continuous gradient
 with Lipschitz constant~$M$, $M \geq m>0$. Consider
 the following inexact gradient method with step size $\gamma \leq \frac{1}{M}$:
 \[
 y^{(k+1)} = y^{(k)} - \gamma\,\left( \nabla \phi(y^{(k)}) + \epsilon^{(k)}\right),\,\,k=0,1,...,
 \]
 with arbitrary initialization $y^{(0)} \in {\mathbb R}^p$
 and $ \epsilon^{(k)} \in {\mathbb R}^p$. Then, for all
 $k=0,1,...$, there holds:
 \begin{equation}
 \label{eqn-inexact-centr-grad}
 \|y^{(k+1)} - y^\star\|\leq (1-\gamma\,m)\,\|y^{(k)} - y^\star\|
 + \gamma\,\|\epsilon^{(k)}\|,
 \end{equation}
 where $y^\star = \mathrm{arg\,min}_{y \in {\mathbb R}^p}\phi(y)$.
 \end{lemma}
 Quantity $ \epsilon^{(k)}$ in~\eqref{eqn-inexact-centr-grad}
  is an inexactness measure that says how far is
  the employed search direction from the
  exact gradient at the iterate~$y^{(k)}$.

\subsection{Intermediate Lemmas and proof of Theorem~\ref{theorem-R-linear-rate}}
 We now carry out convergence proof
 of Theorem~\ref{theorem-R-linear-rate}
 through a sequence of intermediate Lemmas.
 We first split the primal
  error as follows:
  $
  e_x^{(k)} = x^{(k)} - x^{\bullet}$
   $=
   \left(x^{(k)} - \mathbf{1} \otimes \overline{x}^{(k)}\right)$
     $+
     \mathbf{1} \otimes $ $ \left(\overline{x}^{(k)} -
    x^{\star}
    \right)$
    $=:
     \widetilde{x}^{(k)} + \mathbf{1} \otimes \overline{e}_x^{(k)}.
   $
Quantity
$\widetilde{x}^{(k)} =
x^{(k)} - \mathbf{1} \otimes \overline{x}^{(k)}$
 says how mutually
 different are the
 solution estimates~$x_i^{(k)}$'s
 at different nodes;
 quantity
 $\overline{e}_x^{(k)}$
  says how far is
  the global average
  $\overline{x}^{(k)}
  = \frac{1}{N}\sum_{i=1}^N
  x_i^{(k)}$
  from the solution~$x^\star$.
  We
  decompose
  the dual error
  $e_u^{(k)}=u^{(k)} + \nabla F(x^{\bullet})$
    in the following way:
   \[e_u^{(k)} = \widetilde{u}^{(k)}
    + \mathbf{1} \otimes \overline{e}_u^{(k)}.\]
     Here,
     $\overline{e}_u^{(k)} = \frac{1}{N}(\mathbf{1} \otimes I)^\top\,e_u^{(k)} =
   \frac{1}{N}\sum_{i=1}^N [e_u^{(k)}]_i $, and
    $
   \widetilde{u}^{(k)}
    =(I-\mathcal{J})e_u^{(k)}=$
    $e_u^{(k)} - \mathbf{1} \otimes \overline{e}_u^{(k)}
    $. Note that
    $ \overline{e}_u^{(k)} =
    \overline{u}^{(k)} + \frac{1}{N}\sum_{i=1}^N \nabla f_i(x^\star) =
    \overline{u}^{(k)},$
     where
    $\overline{u}^{(k)}
   =\frac{1}{N}\sum_{i=1}^N u_i^{(k)} $.
We will be interested
in deriving bounds on
$\|\mathbf{e_x}\|^{\delta,K}=
\max_{k=0,...,K}\frac{1}{\delta^k}\|e_x^{(k)}\|$,
 for some $\delta \in (0,1)$, and on the analogous
quantities that correspond to other
primal and dual errors that we defined above.
 Specifically, the proof path is as follows.
  First, Lemma~8 shows that
  $\overline{e}_u^{(k)}=0,$ for all $k$,
  which simplifies
  further analysis.
  Our goal is to apply Theorem~\ref{theorem-small-gain}
  with the following identification
  of infinite sequences:
  $\mathbf{a} \rightarrow \mathbf{\widetilde{x}}$,
  and $\mathbf{b} \rightarrow \mathbf{\widetilde{u}}$.
   Then, Lemmas~{9-11} are devoted
   to deriving a bound like in~\eqref{eqn-thm-small-gain-1},
   while Lemma~{12} devises a bound
   that corresponds to~\eqref{eqn-thm-small-gain-2}.
   As shown below, this sequence of Lemmas will
   be sufficient to complete the proof of Theorem~\ref{theorem-R-linear-rate}.

\begin{lemma}
\label{lemma-bar-u-zero}
With algorithm~\eqref{eqn-proposed}--\eqref{eqn-proposed-2}, there holds:
 $\overline{e}_u^{(k)}=\overline{u}^{(k)}=0$, for all $k=0,1,...$
\end{lemma}
\begin{IEEEproof}
Consider~\eqref{eqn-proposed-2}.
Multiplying the equality
from the left by~$\frac{1}{N}(\mathbf{1} \otimes I)^\top$,
using
$(\mathbf{1} \otimes I)^\top \mathcal{L} $
$=
(\mathbf{1}^\top \otimes I) ((I-W)\otimes I)
 $
 $= (\mathbf{1}^\top (I-W))\otimes I = 0$,
 we obtain that:
 \begin{eqnarray}
 \label{eqn-proof-u-bar}
 \overline{u}^{(k+1)}=
 \overline{u}^{(k)}, \,\,k=0,1,...
 \end{eqnarray}
Recall that
$\overline{u}^{(0)}=0,$ by assumption. Thus, the result.
\end{IEEEproof}
\begin{lemma}
\label{lemma-bar-e-to-tilde-x}
Consider algorithm~\eqref{eqn-proposed}--\eqref{eqn-proposed-2},
with $\alpha \leq 1/L$.
 Then, for any $\delta \in \left( 1-\frac{\alpha\,\mu}{2},1\right)$, there holds:
 \begin{equation}
 \label{eqn-lemma-e-bar-tilde-x}
 \|\mathbf{\overline{e}_x}\|^{\delta,K}
 \leq
 \frac{4\, L}{\sqrt{N}\,\mu}\,\|\mathbf{\widetilde{x}}\|^{\delta,K} +
 \frac{2}{\alpha\,\mu}\,\|\overline{e}_x^{(0)}\|.
 \end{equation}
\end{lemma}
\begin{IEEEproof}
Consider~\eqref{eqn-proposed}.
Multiplying the equation from the left
by $\frac{1}{N}(\,\mathbf{1} \otimes I\,)^\top$,
using the fact that
$\frac{1}{N}(\,\mathbf{1} \otimes I\,)^\top\,\mathcal{W} = \frac{1}{N}(\,\mathbf{1} \otimes I\,)^\top$,
 and the fact that $\frac{1}{N}(\,\mathbf{1} \otimes I\,)^\top\,u^{(k)} =
 \overline{u}^{(k)}=0$, for all $k$ (by Lemma~\ref{lemma-bar-u-zero}), we obtain:
 {\small{
 {\allowdisplaybreaks{
 \begin{eqnarray*}
 &\,&\overline{x}^{(k+1)}
 =
 \overline{x}^{(k)}
 - \frac{\alpha}{N}\sum_{i=1}^N \nabla f_i(x_i^{(k)})=\overline{x}^{(k)}
 - \frac{\alpha}{N}\sum_{i=1}^N \nabla f_i(\overline{x}^{(k)}) \\
 &\,&\,\,\,\,\,\,\,\,\,\,\,\,\,\,\,
 - \,\frac{\alpha}{N}\sum_{i=1}^N \left( \nabla f_i(x_i^{(k)}) - \nabla f_i(\overline{x}^{(k)})\right)
 \\
 &\,&\,\,\,\,\,\,\,\,\,\,\,\,\,\,\,=\overline{x}^{(k)}
 - \frac{\alpha}{N} \left(\nabla f(\overline{x}^{(k)}) + \epsilon^{(k)}\right),\,\,\mathrm{where}\\
 &\,& \epsilon^{(k)}=
 \sum_{i=1}^N \left( \nabla f_i(x_i^{(k)}) - \nabla f_i(\overline{x}^{(k)})\right).
 \end{eqnarray*}}}}}
 Now, applying Lemma~\ref{lemma-inexact},
 with $\overline{e}_x^{(k)} = \overline{x}^{(k)}-x^\star$, we obtain:
 \begin{equation}
 \label{eqn-proof-15}
 \|\overline{e}_x^{(k+1)}\|
 \leq
 (1-\alpha\,\mu)\,\|\overline{e}_x^{(k)}\| + \frac{\alpha}{N}\,\|\epsilon^{(k)}\|.
 \end{equation}
 We next upper bound $\frac{\alpha}{N}\,\|\epsilon^{(k)}\|$ as follows:
 {\allowdisplaybreaks{
 \begin{eqnarray}
 \frac{\alpha}{N}\,\|\epsilon^{(k)}\|
  &\leq&
\frac{\alpha}{N} \sum_{i=1}^N \left\| \nabla f_i(x_i^{(k)}) - \nabla f_i(\overline{x}^{(k)})\right\| \nonumber \\
\label{eqn-chain-proof-1-b}
&\,&
\leq
\frac{\alpha}{N} \sum_{i=1}^N \,L\,\left\| x_i^{(k)} - \overline{x}^{(k)} \right\| \\
&\,&
\label{eqn-chain-proof-1-c}
\leq
\frac{\alpha\,L}{\sqrt{N}} \left\| \widetilde{x}^{(k)} \right\|.
 \end{eqnarray}}}
  Inequality \eqref{eqn-chain-proof-1-b}
  is by the Lipschitz continuity of the $\nabla f_i$'s,
  while~\eqref{eqn-chain-proof-1-c} is
  by noting that
  \[
  \sum_{i=1}^N \|x_i^{(k)}-\overline{x}^{(k)}\|
  =  \sum_{i=1}^N \|\widetilde{x}_i^{(k)} \|
  \leq \sqrt{N}\,\|\widetilde{x}^{(k)}\|.
  \]
  Substituting the last bound in \eqref{eqn-proof-15}, we obtain:
 \begin{equation}
 \label{eqn-proof-16}
 \|\overline{e}_x^{(k+1)}\|
 \leq
 (1-\alpha\,\mu )\,\|\overline{e}_x^{(k)}\| +
\frac{\alpha\,L}{\sqrt{N}}\,\|\widetilde{x}^{(k)}\|.
 \end{equation}
Now, applying Lemma~\ref{lemma-small-gain}, we obtain:
 \begin{equation}
 \label{eqn-proof-17}
 \|\mathbf{\overline{e}_x}\|^{\delta,K}
 \leq
 \frac{1}{\delta}(1-\alpha\,\mu )\,\|\mathbf{\overline{e}_x}\|^{\delta,K} +\frac{\alpha\,L}{\sqrt{N}}\frac{1}{\delta}\,\|\mathbf{\widetilde{x}}\|^{\delta,K}
  + \|\overline{e}_x^{(0)}\|.
  %
 \end{equation}
From \eqref{eqn-proof-17}, using $\alpha \leq 1/L$,
one can verify that,
for all $\delta \geq 1 - \frac{\alpha\,\mu}{2}$, there holds:
{\allowdisplaybreaks{
 \begin{eqnarray}
 \|\mathbf{\overline{e}_x}\|^{\delta,K}
 &\leq&
\left( 1-\frac{\alpha\,\mu}{2}\right)\,\|\mathbf{\overline{e}_x}\|^{\delta,K} \nonumber  \\
 \label{eqn-proof-18}
&\,&\,\,2\,\frac{\alpha\,L}{\sqrt{N}}\,\|\mathbf{\widetilde{x}}\|^{\delta,K}
  + \|\overline{e}_x^{(0)}\|.
  %
 \end{eqnarray}}}
The last bound yields the desired result.

%
 %
 %
\end{IEEEproof}
\begin{lemma}
\label{lemma-tilde-x-tobar-e-to-u}
Let
$\alpha < \frac{1-{\sigma}}{3L}$, and $\delta \in \left(1-\frac{\alpha\,\mu}{2},1\right)$.
 Then, there holds:
{\allowdisplaybreaks{
\begin{eqnarray*}
\|\mathbf{\widetilde{x}}\|^{\delta,K}
&\leq&
\frac{12}{5}\,\frac{\alpha\,L\,\sqrt{N}}{1-{\sigma}}
\|\mathbf{\overline{e}_x}\|^{\delta,K} \\
&\,&\,\,+\,
\frac{12}{5}\,\frac{\alpha}{1-{\sigma}}
\|\mathbf{\widetilde{u}}\|^{\delta,K}  +
\frac{2}{1-{\sigma}} \|\widetilde{x}^{(0)}\|.
\end{eqnarray*}}}
\end{lemma}
\begin{IEEEproof}
Consider~\eqref{eqn-proposed}.
Subtracting $x^\bullet = \mathbf{1} \otimes x^\star$
from both sides of
the equation, and noting
that
$\mathcal{W}\,x^\bullet $ $
=
(W \otimes I)(\mathbf{1} \otimes x^\star) $ $=
(W \,\mathbf{1}) \otimes (I \, x^\star) =
\mathbf{1} \otimes x^\star = x^\bullet$,
we obtain:
\begin{eqnarray}
e_x^{(k+1)}
=
\mathcal{W}\,e_x^{(k)} - \alpha \,\left( \nabla F(x^{(k)})+u^{(k)}\right).
\label{eqn-dokaz-0}
\end{eqnarray}
Next, note that
{\allowdisplaybreaks{
\begin{eqnarray}
 &\,&\nabla F(x^{(k)})+u^{(k)} =   \left(\nabla F(x^{(k)}) - \nabla F(x^{\bullet})\right)\nonumber
 \\
 \label{eqn-dokaz-1}
 &\,&\,\, +\,
 \left(\nabla F(x^{\bullet}) + u^{(k)} \right)\\
 \label{eqn-dokaz-2}
 &\,&\,\,=\, \left(\nabla F(x^{(k)}) - \nabla F(x^{\bullet})\right)
 +
 \widetilde{u}^{(k)}.
\end{eqnarray}}}
In~\eqref{eqn-dokaz-2},
we use that
$\nabla F(x^{\bullet}) + u^{(k)} =e_u^{(k)}=\widetilde{u}^{(k)}$ (by Lemma~\ref{lemma-bar-u-zero}).
 Substituting \eqref{eqn-dokaz-2} into \eqref{eqn-dokaz-0}, we get:
 \begin{eqnarray}
e_x^{(k+1)}
=
\mathcal{W}\,e_x^{(k)} - \alpha \,\left(\nabla F(x^{(k)}) - \nabla F(x^{\bullet})\right) - \alpha\,\widetilde{u}^{(k)}
\label{eqn-dokaz-4}
 \end{eqnarray}
We next multiply
\eqref{eqn-dokaz-4}
from the left by $I-\mathcal{J}=(I-J) \otimes I$, and
we use that
$[\,(I-J) \otimes I\,]e_x^{(k)} = \widetilde{x}^{(k)}$.
Noting that
\begin{eqnarray*}
\left[ \,(I-J) \otimes I\,\right]\,\mathcal{W}
&=& \left[ \,(I-J) \otimes I\,\right]\,[\,W \otimes I\,] \\
&\,&=
  [\,(I-J)W\,]\otimes I \\
  &\,&
=\widetilde{W} \otimes I =\widetilde{\mathcal{W}},
\end{eqnarray*}
 and $(I -\mathcal{J})\widetilde{u}^{(k)} = \widetilde{u}^{(k)}$, we get:
 \begin{eqnarray}
\widetilde{x}^{(k+1)}
&=&
\widetilde{\mathcal{W}}\,\widetilde{x}^{(k)}  - \alpha \,(I-\mathcal{J})\,\left(\nabla F(x^{(k)})\right. \nonumber\\
&-& \left. \nabla F(x^{\bullet})\right) - \alpha\,\widetilde{u}^{(k)}.
\label{eqn-tilde-x-norm-2}
 \end{eqnarray}
 %
 Next, use the decomposition
 $x^{(k)} - x^\bullet=e_x^{(k)} = \widetilde{x}^{(k)} +
 \mathbf{1} \otimes \overline{e}_x^{(k)}$,
 and Lipschitz continuity of $\nabla F$,
 to note that:
 \begin{equation}
 \|\nabla F(x^{(k)}) - \nabla F(x^{\bullet})\|
 \leq L\,\|\widetilde{x}^{(k)}\|
 +L\,\sqrt{N}\,\|\overline{e}_x^{(k)}\|.
 \end{equation}
 Using the latter bound and taking the 2-norm in \eqref{eqn-tilde-x-norm-2},
  while using its sub-additive
  and sub-multiplicative properties, we get:
  \begin{equation}
  \|\widetilde{x}^{(k+1)}\|
  \leq
  \left( {\sigma}+\alpha\,L\right)\,
  \|\widetilde{x}^{(k)}\|
  + \alpha\,L\,\sqrt{N}\,\|\overline{e}_x^{(k)}\|
  +\alpha\,\|\widetilde{u}^{(k)}\|.
  \label{eqn-proof-10}
  \end{equation}
  Now, similarly to Lemma~\ref{lemma-small-gain}, it is easy to see that
  the last equation implies:
  \begin{eqnarray}
  &\,&\|\mathbf{\widetilde{x}}\|^{\delta,K}
  \leq
  \frac{1}{\delta}\left( {\sigma}+\alpha\,L
  \right)\,\|\mathbf{\widetilde{x}}\|^{\delta,K}\nonumber\\
  &\,&\,\,\,\,+ \frac{1}{\delta}\alpha\,L\,\sqrt{N}\,\|\mathbf{\overline{e}_x}\|^{\delta,K}
  + \frac{\alpha}{\delta} \|\mathbf{\widetilde{u}}\|^{\delta,K}
  + \|\widetilde{x}^{(0)}\|.
  \label{eqn-proof-111}
  \end{eqnarray}
Next, note that for
$\delta \in \left( 1-\frac{\alpha\,\mu}{2},1\right)$,
and
$\alpha < \frac{1-{\sigma}}{3L}$, there holds:
\begin{equation}
\frac{1}{\delta}\left( {\sigma}+\alpha\,L \right)
  < \frac{{\sigma}+1}{2}.
\end{equation}
Also, as $\alpha < \frac{1-{\sigma}}{3L} < \frac{1}{3L}$, we have that
$1/\delta \leq 6/5$.
 Substituting
 the last two bounds in \eqref{eqn-proof-111},
 we obtain:
\begin{eqnarray*}
  \|\mathbf{\widetilde{x}}\|^{\delta,K}
  &\leq&
  \frac{1+{\sigma}}{2}\,
  \|\mathbf{\widetilde{x}}\|^{\delta,K}
  + \frac{6}{5}\alpha\,L\,\sqrt{N}\,\|\mathbf{\overline{e}_x}\|^{\delta,K} \\
  &\,&\,\,+ \,\frac{6\,\alpha}{5} \|\mathbf{\widetilde{u}}\|^{\delta,K}
  + \|\widetilde{x}^{(0)}\|.
  \end{eqnarray*}
  Rearranging
  terms in the last equality,
  the desired result follows.
\end{IEEEproof}
\begin{lemma}
\label{lemma-tilde-x-to-solo-u}
Let $\alpha < \frac{5}{96}\,\frac{(1-{\sigma})\mu}{L^2}$, and $\delta \in \left(1-\frac{\alpha\,\mu}{2},1\right)$.
 Then, there holds:
\begin{eqnarray}
&\,&\|\mathbf{\widetilde{x}}\|^{\delta,K}
\leq
\frac{24}{5}\,\frac{\alpha}{1-{\sigma}}
\|\mathbf{\widetilde{u}}\|^{\delta,K}  +
\frac{4}{1-{\sigma}} \|\widetilde{x}^{(0)}\|\\
&\,&\,\,\,\,\,\,\,\,\,\,\,\,\,\,\,+\,\,
\frac{48}{5}\,\frac{L\,\sqrt{N}}{\mu\,(1-{\sigma})} \|\overline{e}_x^{(0)}\|. \nonumber
\end{eqnarray}
\end{lemma}
\begin{IEEEproof}
We combine Lemmas \ref{lemma-bar-e-to-tilde-x} and \ref{lemma-tilde-x-tobar-e-to-u}, to obtain:
\begin{eqnarray}
\|\mathbf{\widetilde{x}}\|^{\delta,K}
&\leq&
\frac{48}{5}\,\frac{\alpha\,L^2}{(1-{\sigma})\,\mu}
\,\|\mathbf{\widetilde{x}}\|^{\delta,K}
+
\frac{24}{5}\,\frac{\sqrt{N}\,L}{(1-{\sigma})\,\mu}\,\|\overline{e}_x^{(0)}\| \nonumber
\\
&+&
\label{eqn-proof-1000}
\frac{12}{5}\,\frac{\alpha}{1-{\sigma}}\,\|\mathbf{\widetilde{u}}\|^{\delta,K}
+
\frac{2}{1-{\sigma}}\,\|\widetilde{x}^{(0)}\|.
\end{eqnarray}
Next, note that,
for $\alpha<\frac{5}{96}\,\frac{(1-{\sigma})\mu}{L^2}$,
we have that
 $\frac{12\,\alpha\,L^2}{(1-{\sigma})\,\mu}<1/2$.
 Substituting the latter bound in
 \eqref{eqn-proof-1000}
 and manipulating the terms, the
 desired result follows.
\end{IEEEproof}
\begin{lemma}
\label{lemma-u-to-x-tilde}
Let $\delta \in \left( \frac{1+{\sigma}}{2},\,1 \right)$,
and recall $L^{\prime}=\left( L^2+b^2 - 2\,b\,\mu\right)^{1/2}$.
  Then, the following holds:
 \begin{eqnarray}
 \|\mathbf{u}\|^{\delta,K}
 &\leq&
 \frac{40\,L^\prime\,L}{\mu\,(1-{\sigma})}
 \,\|\mathbf{\widetilde{x}}\|^{\delta,K}
 +
 \frac{2}{1-{\sigma}}\|\widetilde{u}^{(0)}\|\nonumber\\
 &\,&\,\,+\,
 \frac{16}{1-{\sigma}}\,\frac{L^\prime\,\sqrt{N}}{\alpha\,\mu}\,\| {\overline{e}_x}^{(0)}\|.\nonumber
 \end{eqnarray}
\end{lemma}
\begin{IEEEproof}
Consider \eqref{eqn-proposed-2}. Adding
$\nabla F(x^\bullet)$ to both sides of the equality,
and recalling that $e_u^{(k)} = u^{(k)}+\nabla F(x^\bullet) = \widetilde{u}^{(k)}$, we obtain:
\begin{eqnarray}
\widetilde{u}^{(k+1)}
&=&
\widetilde{u}^{(k)} - \mathcal{L}\,(\, \nabla F(x^{(k)}) - \nabla F(x^{\bullet}) + u^{(k)} \nonumber
\\&\,&\,\,+\, \nabla F(x^{\bullet}) - b\, (x^{(k)}-x^\bullet)\,)
\label{eqn-proposed-2-new}\\
&\,&=\,\,\mathcal{W}\,\widetilde{u}^{(k)} - \mathcal{L}\,(\, \nabla F(x^{(k)}) - \nabla F(x^{\bullet})\,) \nonumber \\
&\,&\,\, +\,
b\,\mathcal{L}\, (x^{(k)}-x^\bullet)
\nonumber \\
&\,&=\,\,
\widetilde{\mathcal{W}}\,\widetilde{u}^{(k)} - \mathcal{L}\,(\, \nabla F(x^{(k)}) - \nabla F(x^{\bullet})\,)
\nonumber \\
&\,&\,\,+\,
b\,\mathcal{L}\, (x^{(k)}-x^\bullet).
\label{eqn-proposed-2-new-3}
\end{eqnarray}
Here, \eqref{eqn-proposed-2-new} uses
the fact that $\mathcal{L}\,x^\bullet=0$, while
 \eqref{eqn-proposed-2-new-3}
 holds because $\widetilde{\mathcal{W}}\,
 \widetilde{u}^{(k)}
 =
 ({\mathcal{W}} - \mathcal{J})\,
 \widetilde{u}^{(k)}
 $
 $=
 {\mathcal{W}}\,
 \widetilde{u}^{(k)}
 $, as
 $\mathcal{J} \widetilde{u}^{(k)}
 =
 \mathcal{J} (I - \mathcal{J}) {e_u}^{(k)}
 $
  $
  = (\mathcal{J} - \mathcal{J}) {e_u}^{(k)} = 0
  $.
  We next upper bound
  the term
  $\left(
  \nabla F(x^{(k)})
  -\nabla F(x^{\bullet}) - b\,(x^{(k)}-x^{\bullet})\,
  \right)$ as follows:
   {\allowdisplaybreaks{
   \begin{eqnarray*}
   &\,&   \|\nabla F(x^{(k)}) -
   \nabla F(x^{\bullet})
   - b \,(x^{(k)}-x^{\bullet})  \|^2\\
   &\,&\,\,=
   \|
   \nabla F(x^{(k)}) -
   \nabla F(x^{\bullet})\|^2
   +
   b^2\,\|x^{(k)}-x^{\bullet}\|^2\\
   &\,&\,\,
   -2\,b\,\left( \nabla F(x^{(k)}) -
   \nabla F(x^{\bullet})\right)^\top \left( x^{(k)}-x^{\bullet}\right) \\
   &\,&\,\,
   \leq
   L^2\,\|x^{(k)}-x^{\bullet}\|^2
   +
   b^2\,\|x^{(k)}-x^{\bullet}\|^2\\
   &\,&\,\,-\,
   2\,b\,\mu\,\|x^{(k)}-x^{\bullet}\|^2,
   \end{eqnarray*}}}
  where the last
  inequality
  holds by the Lipschitz
  continuity
  of $\nabla F$,
  and by the strong monotonicity
  of~$\nabla F$:
  \[
  \left( \nabla F(x) -
   \nabla F(y)\right)^\top \left( x-y\right)
   \geq \mu\,\|x-y\|^2,\,\,\,\,\mathrm{for\,\,all}\,\,\,\,x,y \in {\mathbb R}^d.
  \]
  Hence,
   we have that:
  \[
  \|\nabla F(x^{(k)})
  -\nabla F(x^{\bullet}) - b\,(x^{(k)}-x^{\bullet})\|
  \leq
  L^\prime\,\|x^{(k)}-x^{\bullet}\|.
  \]
  Next, taking the norm
  in~\eqref{eqn-proposed-2-new-3}, using $\|\mathcal{L}\| \leq 2$, exploiting
  its sub-additive and sub-multiplicative properties,
  and using the Lipschitz continuity of $\nabla F$, we obtain:
\begin{eqnarray}
\label{eqn-proposed-2-new-4}
\|\widetilde{u}^{(k+1)}\| \leq
 {\sigma} \,\|\widetilde{u}^{(k)}\| + 2\,L^\prime\,\, \|e_x^{(k)}\| .
\end{eqnarray}
Decomposing $e_x^{(k)}=
\widetilde{x}^{(k)}+\mathbf{1} \otimes
\overline{e}_x^{(k)}$, and applying Lemma~\ref{lemma-small-gain},
we obtain:
\begin{eqnarray*}
\|\mathbf{\widetilde{u}}\|^{\delta,K}
&\leq&
\frac{{\sigma}}{\delta}\,\|\mathbf{\widetilde{u}}\|^{\delta,K}
+
\frac{2\,L^\prime}{\delta}\,\|\mathbf{\widetilde{x}}\|^{\delta,K}\\
&+& \frac{2 \,L^\prime \,\sqrt{N}}{\delta}
 \|\mathbf{\overline{e}_x}\|^{\delta,K}
 +\|\widetilde{u}^{(0)}\|.
 \end{eqnarray*}
 Next, applying Lemma~\ref{lemma-bar-e-to-tilde-x}, we get:
 {\allowdisplaybreaks{
\begin{eqnarray*}
\|\mathbf{\widetilde{u}}\|^{\delta,K}
&\leq&
\frac{{\sigma}}{\delta}\,\|\mathbf{\widetilde{u}}\|^{\delta,K}
+
\left(\frac{2\,L^\prime}{\delta}\right)\,\left(1+\frac{4\,L}{\mu}\right)\,\|\mathbf{\widetilde{x}}\|^{\delta,K}\\
&+&\frac{4\,\sqrt{N}\,L^\prime}{\alpha\,\mu\,\delta} \,\|\overline{e}_x^{(0)}\|
 +\|\widetilde{u}^{(0)}\|.
 \end{eqnarray*}}}
 From now on, assume that
 $\delta \in \left( \frac{1+{\sigma}}{2},\,1\right)$. Then,
 it is easy to see that there holds:
 $
\frac{{\sigma}}{\delta} < \frac{1+{\sigma}}{2}.
 $
Also, note that $\frac{1}{\delta}\leq 2$.
Thus, we have:
\begin{eqnarray*}
\|\mathbf{\widetilde{u}}\|^{\delta,K}
&\leq&
\frac{1+{\sigma}}{2}\,\|\mathbf{\widetilde{u}}\|^{\delta,K}
+
\frac{20\,L^\prime\,L}{\mu}\,\|\mathbf{\widetilde{x}}\|^{\delta,K}\\
&+&\frac{8 \,L^\prime \,\sqrt{N}}{\alpha\,\mu}
 \|{\overline{e}_x}^{(0)}\|^{\delta,K}
 +\|\widetilde{u}^{(0)}\|.
 \end{eqnarray*}
After rearranging expressions, the desired result follows.
\end{IEEEproof}
We are now ready to prove Theorem~\ref{theorem-R-linear-rate}.
\begin{IEEEproof}[Proof of Theorem~\ref{theorem-R-linear-rate}]
We apply Theorem~\ref{theorem-small-gain}
with the identification $\mathbf{a} \rightarrow \mathbf{\widetilde{x}}$
and $\mathbf{b} \rightarrow \mathbf{\widetilde{u}}$,
by utilizing Lemma~11 and Lemma~12.
 Assume the algorithm
parameters obey the
conditions of Lemmas 11 and 12,
namely that:
\[
\alpha<\frac{5}{96}\,\frac{\mu\,(1-{\sigma})}{L^2}\,\,\,\,\,\,\mathrm{and}\,\,\,\,\,\,
\delta \in \left( \max\{\frac{1+{\sigma}}{2},\,1-\frac{\alpha\,\mu}{2}\},1\right).
\]
Then, by the two Lemmas,
the product of gains equals:
 $
\gamma_1\,\gamma_2 = $ $ \frac{6 \alpha}{1-{\sigma}}\,\frac{40\,L^\prime\,L}{\mu\,(1-{\sigma})}
 $
We need that $\gamma_1\,\gamma_2<1$ in order
for~\eqref{eqn-small-gain-claim} to hold.
Therefore,
when
 $
\alpha <\min\{\frac{\mu\,(1-{\sigma})^2}{192\,L^\prime\,L},\frac{\mu(1-\sigma)}{19\,L^2}\},
 $
 we have that
 $
\|\widetilde{x}\|^{\delta} \leq C < \infty,
 $
for a constant $C \in (0, \infty)$.
Note also that, by Lemma~\ref{lemma-bar-e-to-tilde-x},
we have:
 \begin{equation}
 \label{eqn-lemma-e-bar-tilde-x-new}
 \|\mathbf{\overline{e}_x}\|^{\delta}
 \leq
 \frac{4\, L}{\sqrt{N}\,\mu}\,\|\mathbf{\widetilde{x}}\|^{\delta} +
 \frac{2}{\alpha\,\mu}\,\|\overline{e}_x^{(0)}\|.
 \end{equation}
 As~$e_x^{(k)} = \widetilde{x}^{(k)} + \mathbf{1}\otimes \overline{e}_x^{(k)}$,
 we have:
  $
 \|\mathbf{e_x}\|^{\delta} \leq $
 $\left(1+\frac{4\, L}{\sqrt{N}\,\mu}\right)\,\|\mathbf{\widetilde{x}}\|^{\delta} $
 $+
 \frac{2}{\alpha\,\mu}\,\|\overline{e}_x^{(0)}\|$
  $=
 \left(1+\frac{4\, L}{\sqrt{N}\,\mu}\right)\,C+$
 $\frac{4}{\alpha\,\mu}\,\|\overline{e}_x^{(0)}\| $$=: C^\prime <+\infty.$
 Therefore,
  $
 \|e_x^{(k)}\| \leq C^\prime\,\delta^k,\,\,\mathrm{for\,\,all}\,\,k,
 $
 for any $\delta \in \left( \max\{\frac{1+{\sigma}}{2},\,1-\frac{\alpha\,\mu}{2}\},1\right)$, as desired.
\end{IEEEproof}


\section{Simulations}
This Section  provides a simulation example on
learning a linear classifier via minimization of the $\ell_2$-regularized logistic loss.
Simulations confirm the insights gained through the
theoretical analysis and demonstrate that
the proposed generalized method improves convergence
speed over the existing methods~\cite{WotaoYinExtra,Harnessing}.

The simulation setup is as follows. We consider distributed learning of
 a linear classifier via the $\ell_2$-regularized logistic loss, e.g.,~\cite{BoydADMoM}.
 Each node $i$ has $J$ data samples
 $\{a_{ij}, b_{ij}\}_{j=1}^{J}$. Here, $a_{ij} \in {\mathbb R}^{d-1}$, $d \geq 2$, is a feature vector, and
$b_{ij} \in \{-1,+1\}$ is its class label.
The goal is to learn a vector $x=(x_1^\top,x_0)^\top$,
$x_1 \in {\mathbb R}^{d-1}$, and
$x_0 \in {\mathbb R}$, $d \geq 1$,
such that the total $\ell_2$-regularized surrogate loss $\sum_{i=1}^N f_i(x)$ is minimized, where,
for $i=1,...,N$, we have:
\begin{equation*}
f_i(x) =
\mathrm{ln}\left( 1+\mathrm{exp}\left( -b_{ij}\,(\,a_{ij}^\top x_1 + x_0\,)\,\right)\right) + \frac{1}{2}\mathcal{R} \|x\|^2.
\end{equation*}
 Here, $\mathcal{R} $ is a positive regularization parameter.
 We can take the strong convexity constant
  as $\mu = {\mathcal{R}}$, while a Lipschitz
 constant $L$ can be taken as $\frac{1}{4 N} \|\sum_{i=1}^N \sum_{j=1}^J c_{ij}\,c_{ij}^\top\| + {\mathcal R}$,
 where $c_{ij} = (b_{ij}\,a_{ij}^\top, b_{ij})^\top$.

With all experiments, we test the algorithms on a connected network with $N=30$ nodes and $123$ links,
 generated as a realization of the random geometric graph model
 with communication radius~$\sqrt{\mathrm{ln}(N)/N}$.

We generate data and set the algorithm parameters as follows.
 Each node $i$ has $J=2$ data points whose dimension is $d-1=5.$
 The $a_{ij}$'s are generated independently over $i$ and $j$; each entry of $a_{ij}$ is drawn independently
from the standard normal distribution.
We generate the ``true'' vector $x^\star=((x_1^\star)^\top, x_0^\star)^\top$
by drawing its entries independently from standard normal distribution.
 The class labels are generated as $b_{ij}=\mathrm{sign} \left( (x^\star_1)^\top a_{ij}+x^\star_0+\epsilon_{ij}\right)$, where
$\epsilon_{ij}$'s are drawn independently from normal distribution with zero mean and
variance~$0.4$.  We set the regularization parameter as $\mathcal{R}=0.03$.

With all algorithms, we initialize $x_i^{(0)}$ to zero for all $i=1,...,N$.
The auxiliary variables for each
algorithm are initialized as described in Subsection~{II-B}.
 Further, the weight matrix is
as follows:
 $W_{ij}=\frac{1}{2\,\left(\max\{\mathrm{deg}(i),\mathrm{deg}(j)\}+1\right)}$,
  for $i \neq j$,
  $\{i,j\} \in E$;
  $W_{ij}=0$,
  for $i \neq j$,
  $\{i,j\} \notin E$; and
  $W_{ii}=1-\sum_{j \neq i}W_{ij}$, for $i=1,...,N$.
   Here,
  $\mathrm{deg}(i)$
   is the number of neighbors of node~$i$~(excluding~$i$).

As an error metric, we use
   the following quantity:
    $
   \frac{1}{N} \sum_{i=1}^N \frac{\|x_i^{(k)} - x^\star\|}{\|x^\star\|}, \,\,x^\star \neq 0,
    $
   that we refer to as the relative error.
   Quantity $x^{\star}$ is obtained
   numerically beforehand by a
   centralized Nesterov gradient method~\cite{Nesterov-Gradient}.
   %
   %

We compare four methods:
 the method in~\cite{Harnessing},
 that we refer to here as ``harnessing'';
 the Extra method in~\cite{WotaoYinExtra};
  the proposed method~\eqref{eqn-proposed}--\eqref{eqn-proposed-2}
 with $\mathcal{B}=\frac{L+\mu}{2}I$ --
 that we refer to as the modified ``harnessing'';
 and the proposed method~\eqref{eqn-proposed}--\eqref{eqn-proposed-2}
 with $\mathcal{B}=L\,\mathcal{W}$ --
 that we refer to as the modified Extra.

Figure~1 plots the relative
 error versus number of iterations~$k$
 for the four methods,
 for different values
 of step sizes:
 the top Figure: $\alpha=1/(3L)$;
 middle: $\alpha=1/(9L)$;
 and bottom: $\alpha=1/(15 L)$.
 First, on the top Figure,
 we can see that
 the proposed modifications
 yield improvements
 in the convergence speed
 over the respective original methods in~\cite{Harnessing} and~\cite{WotaoYinExtra}.
  While the improvement is not very large for~\cite{WotaoYinExtra}, it is
  quite significant for the method in~\cite{Harnessing}.

   As the step size decreases (the middle and bottom Figures),
   we can see that the gain of the proposed method is reduced,
    and the four methods tend to behave
    mutually very similarly.
    Next, while for the large step size (the top Figure)
    Extra~\cite{WotaoYinExtra} performs better than
     the method in~\cite{Harnessing},
     for the small step size (bottom Figure)
      the performance of the two methods
      is reversed, as predicted by our theoretical considerations. (Though the difference between
      the methods is quite small for the small step size.)

\begin{figure}[thpb]
      \centering
      \includegraphics[angle=-90,origin=c,height=2.9 in,width=3.2 in]{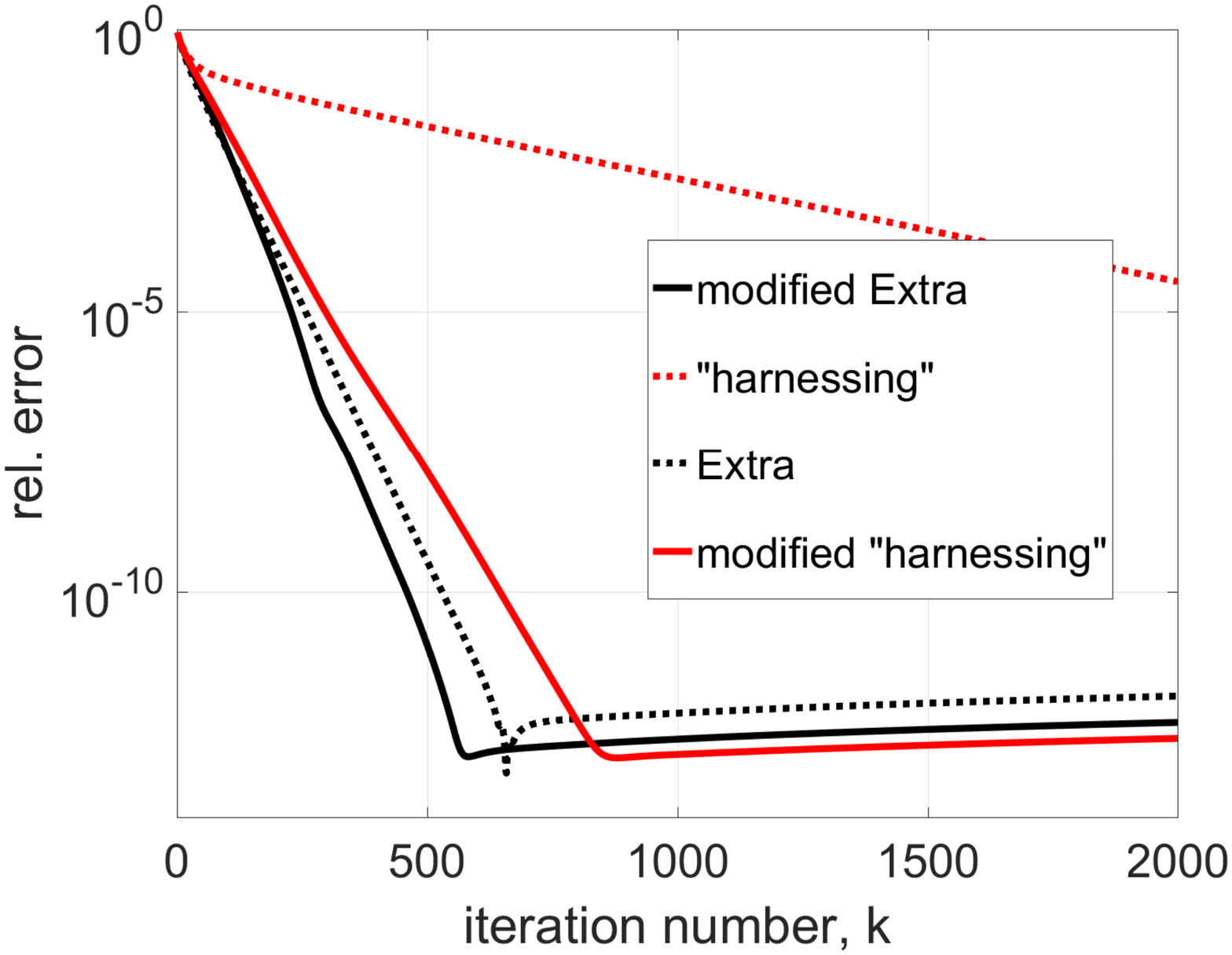}
      \includegraphics[angle=-90,origin=c,height=2.9 in,width=3.2 in]{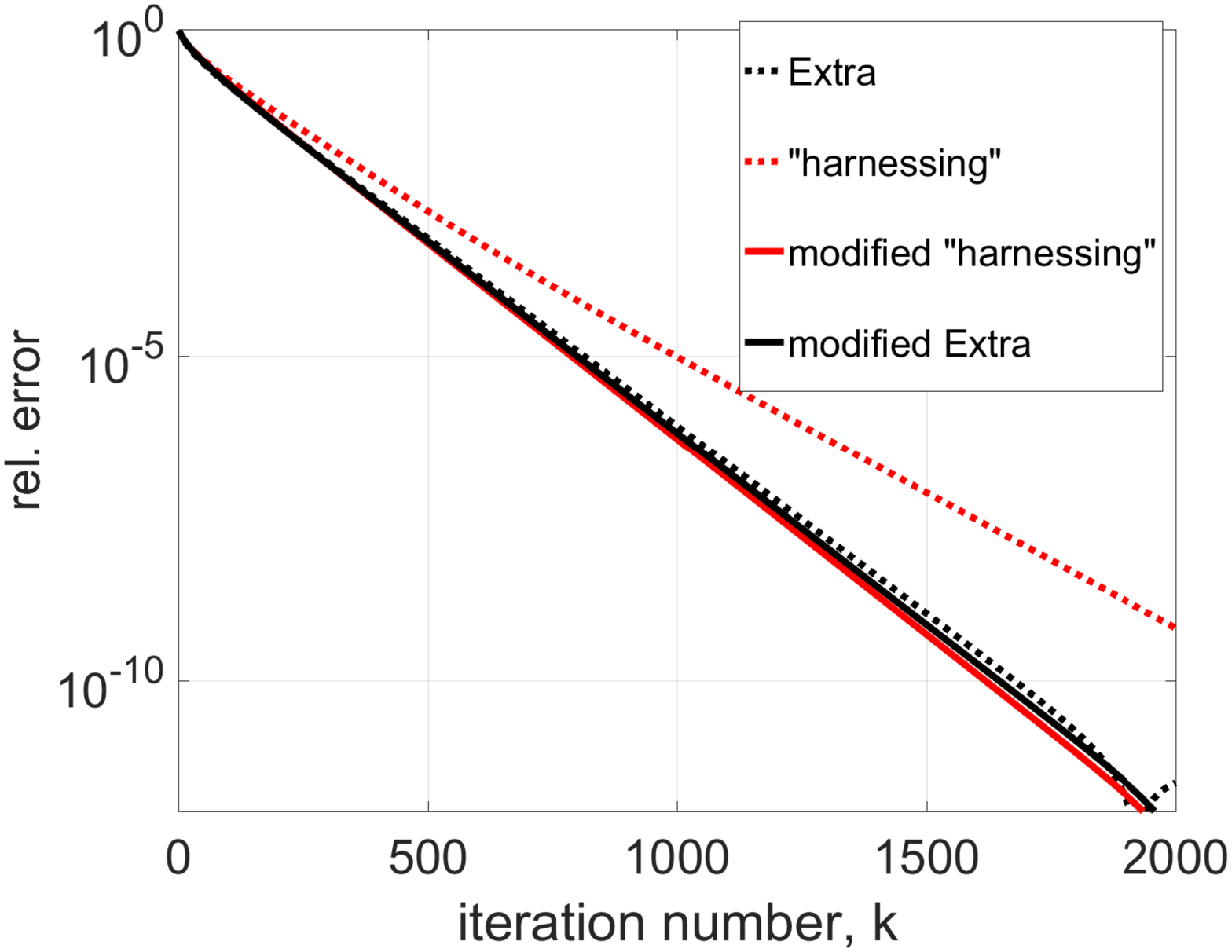}
      \includegraphics[angle=-90,origin=c,height=2.9 in,width=3.2 in]{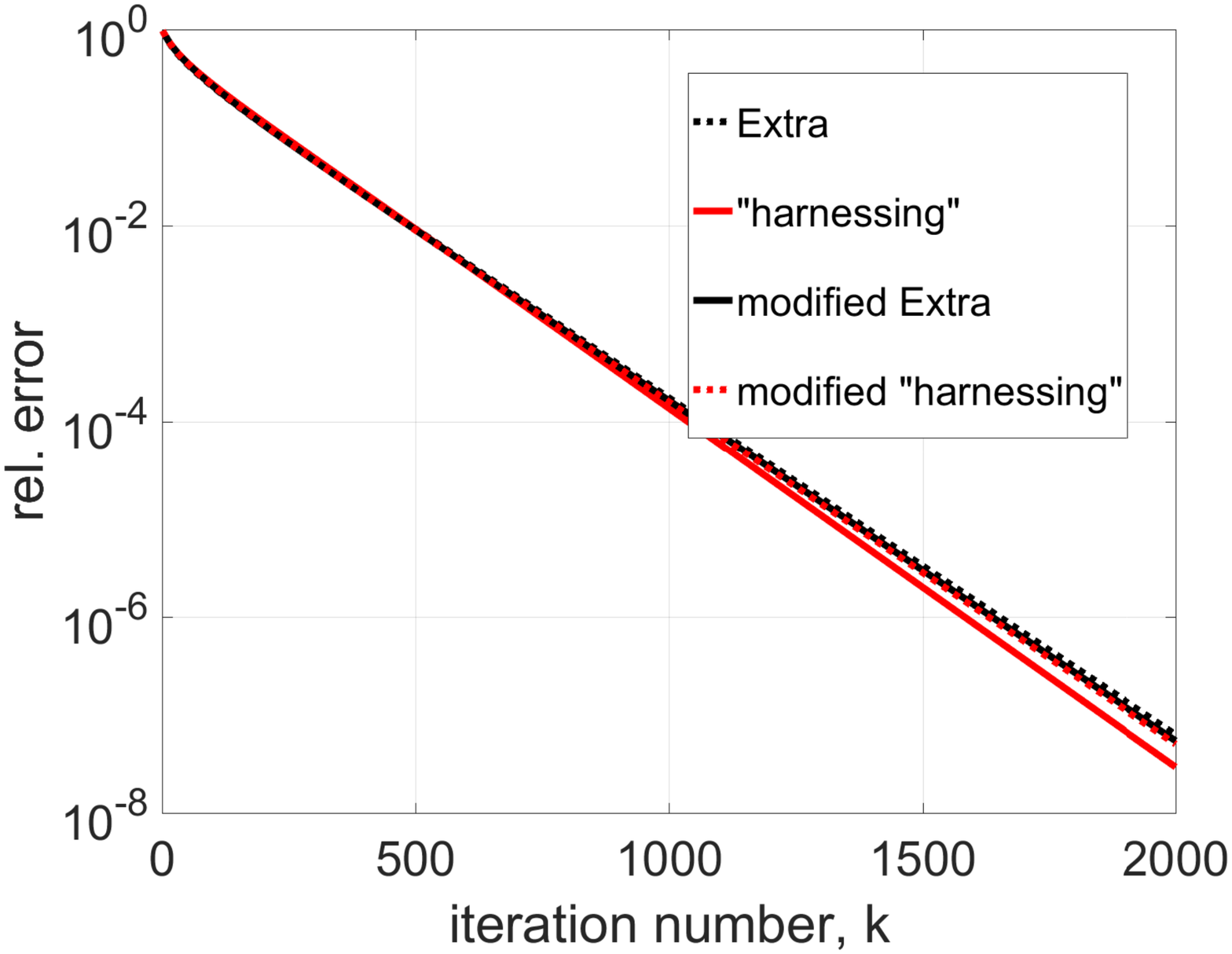}
      \caption{Relative error versus number of iterations~$k$,
for three different values of step-size~$\alpha$: Top:
$\alpha =\frac{1}{3\,L}$; middle: $\alpha=
\frac{1}{9\,L}$; and bottom: $\alpha=\frac{1}{15\,L}.$}
      \label{Figure_1}
\end{figure}
Figure~2 repeats the experiment for
a $100$-node, $561$-link, connected network (generated also as an instance
of a random geometric graph model with radius~$\sqrt{\mathrm{ln}(N)/N}$),
 and for step-sizes
$\alpha=1/(6L)$ (top Figure);
$\alpha=1/(18L)$ (middle);
and $\alpha=1/(54L)$ (bottom).
We can see that a similar
 behavior of the four methods
 can be observed here, as well.

\begin{figure}[thpb]
      \centering
      \includegraphics[angle=-90,origin=c,height=2.9 in,width=3.2 in]{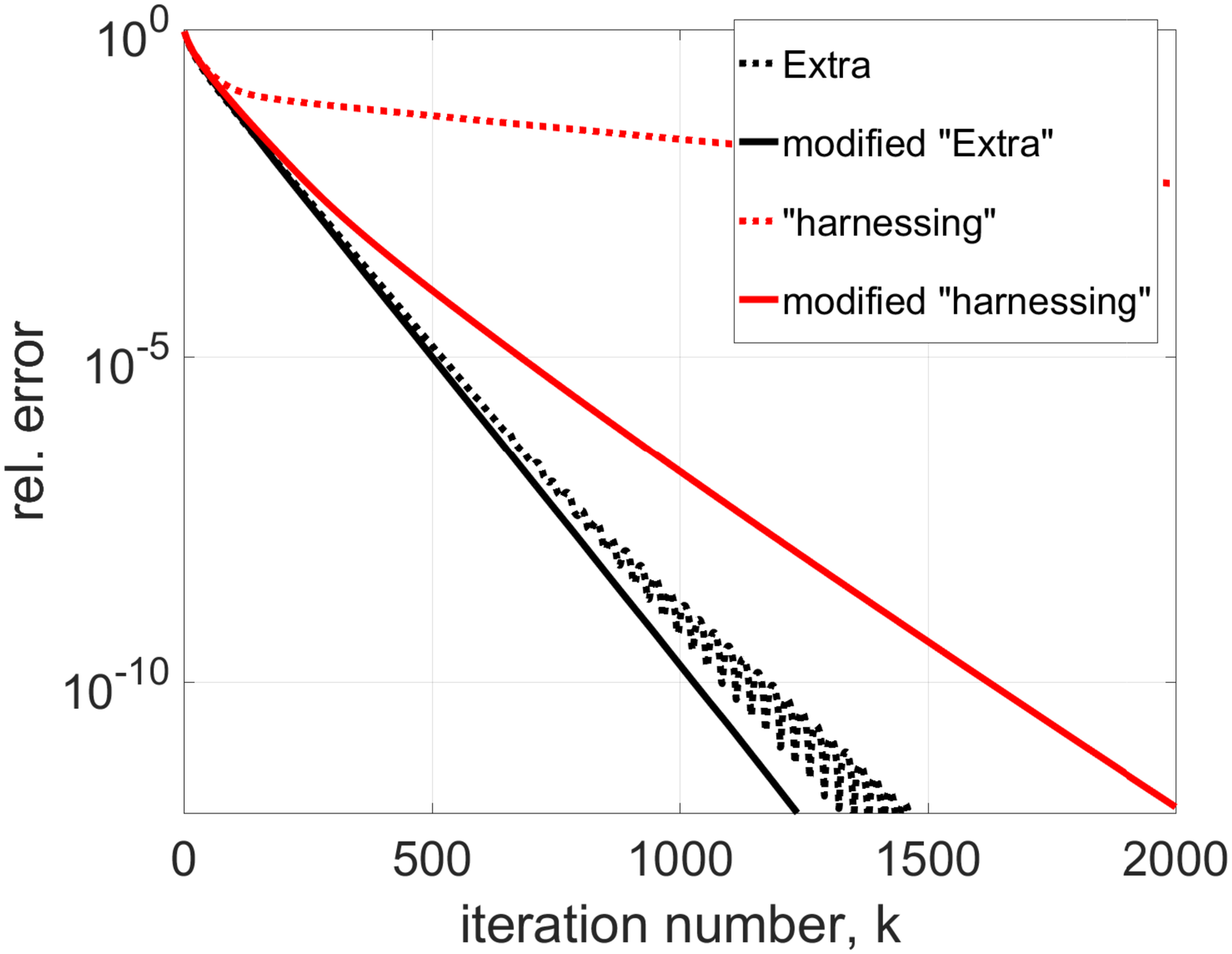}
      \includegraphics[angle=-90,origin=c,height=2.9 in,width=3.2 in]{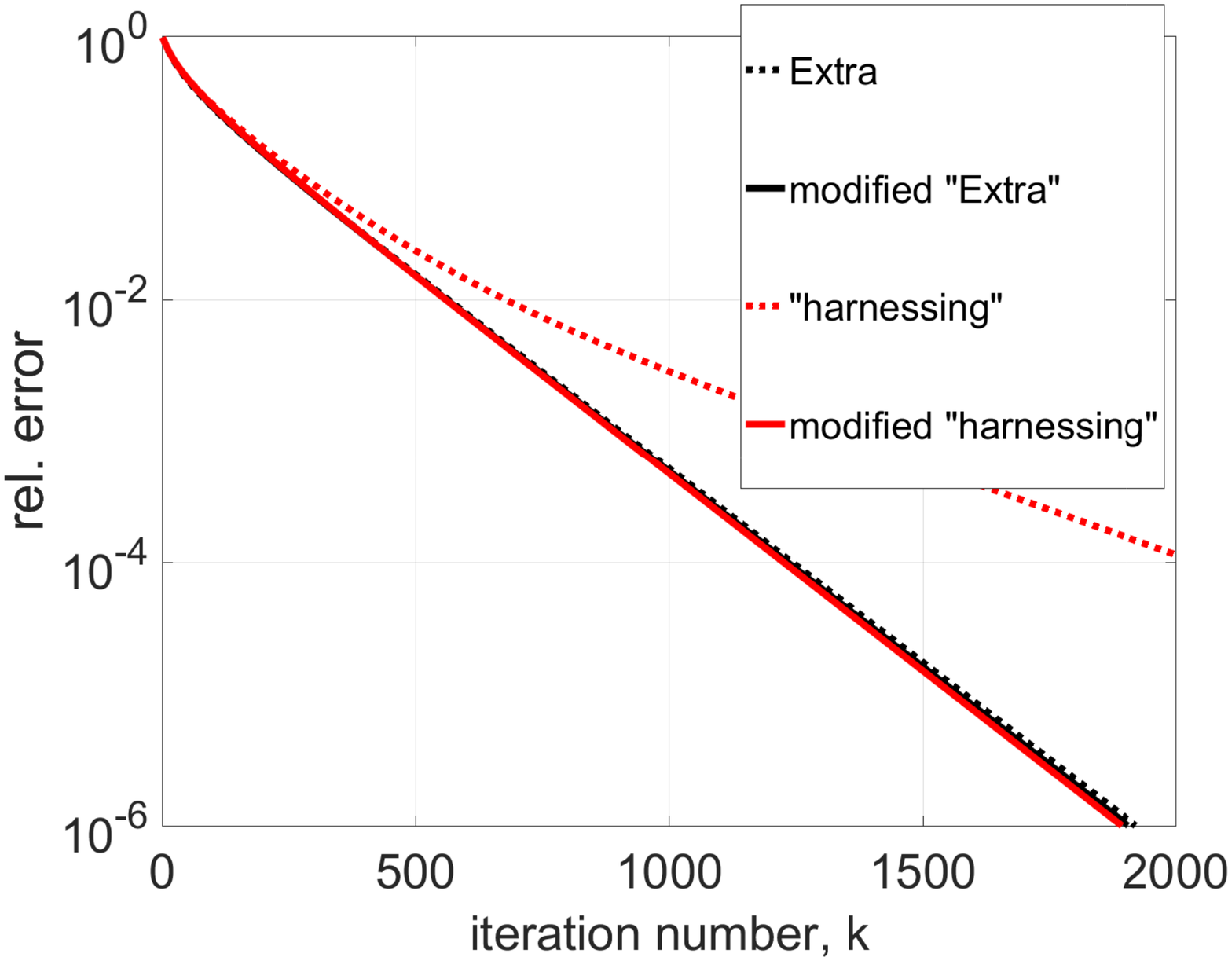}
      \includegraphics[angle=-90,origin=c,height=2.9 in,width=3.2 in]{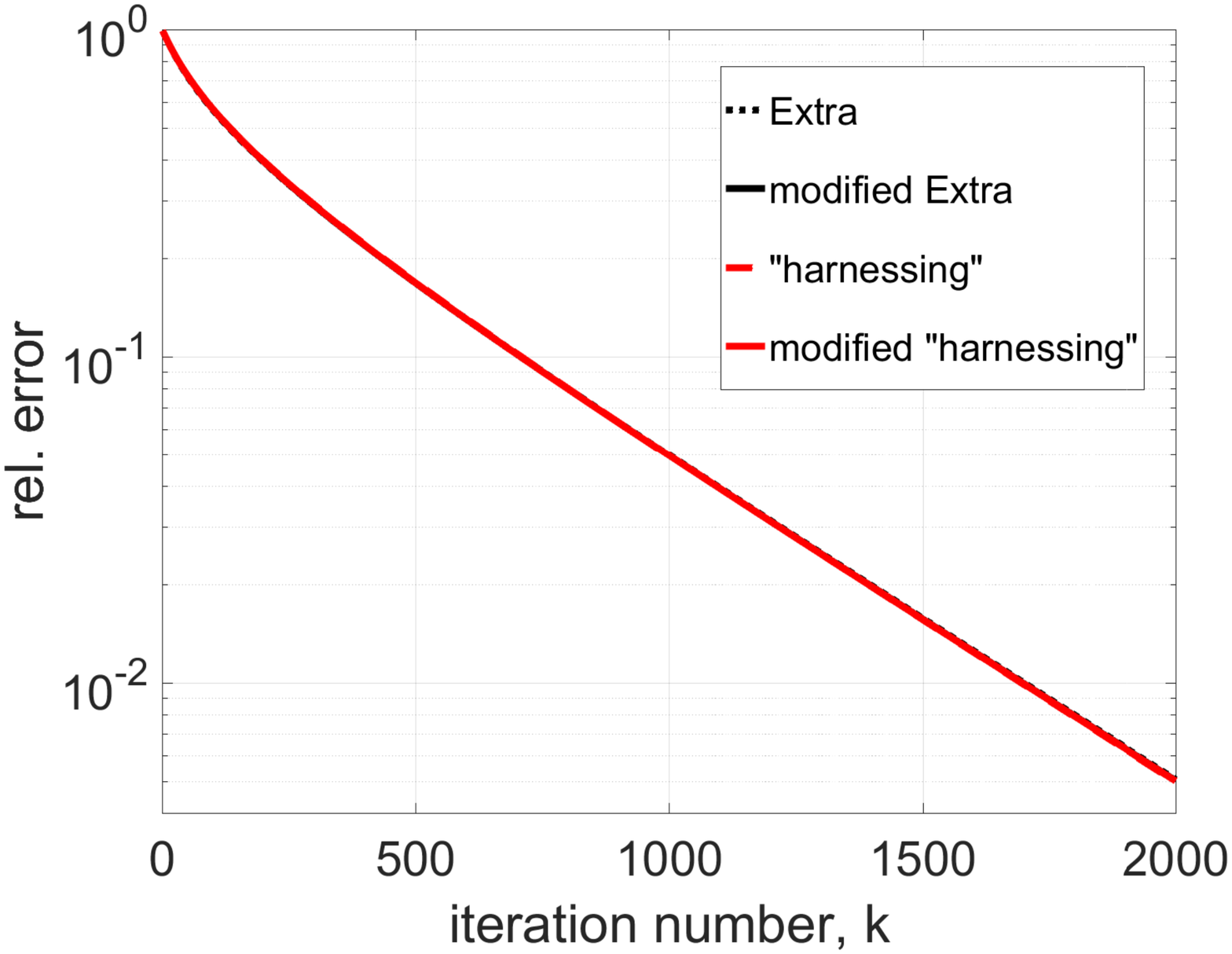}
      \caption{Relative error versus number of iterations~$k$,
for three different values of step-size~$\alpha$: Top:
$\alpha =\frac{1}{6\,L}$; middle: $\alpha=
\frac{1}{18\,L}$; and bottom: $\alpha=\frac{1}{54\,L}.$}
      \label{Figure_1}
\end{figure}

\section{Conclusion}
We considered exact first order
methods for distributed optimization problems
where~$N$ nodes collaboratively minimize
the aggregate sum of their local convex costs.
 Specifically, we unified, generalized, and
 improved convergence speed of the existing methods, e.g.,~\cite{WotaoYinExtra,Harnessing}.
   While it was known that the method in~\cite{WotaoYinExtra}
   is equivalent to a primal-dual gradient-like method,
   we show here that this is true also with~\cite{Harnessing},
   where the corresponding primal-dual update rule
   incorporates a weighted past dual gradient term, in addition to
   the current dual gradient term.
   We then generalize the method
   by proposing an optimized, easy-to-tune weighting
   of the past dual gradient term,
   and show both theoretically and
   by simulation that the modification yields
   significant improvements in convergence speed.
   We establish for the proposed exact method
   global R-linear convergence rate, assuming
   strongly convex local costs with
   Lipschitz continuous gradients
    and static networks. Possible future
    work directions include extensions to uncoordinated step-sizes across nodes
    and time varying, random, and directed networks.
    Specifically, with directed networks,
    it would be interesting to extend the methods
    to work with singly stochastic matrices, i.e.,
    to relax the requirement on the double stochasticity, as
    doubly stochastic methods become undesirable when
dealing with time-varying or directed graph topologies.

    .

\vspace{-3mm}

\section*{Appendix}
\subsection{Proof of Lemma~\ref{lemma-equivalence}}
We first prove part~(b), i.e., we set $\mathcal{B}=\frac{1}{\alpha}\mathcal{W}$. We must  show
that \eqref{eqn-proposed}--\eqref{eqn-proposed-2} is
equivalent to \eqref{eqn-extra}--\eqref{eqn-extra-222}.
 It suffices to show
that~\eqref{eqn-proposed}--\eqref{eqn-proposed-2} is
equivalent to~\eqref{eqn-saddle-extra}--\eqref{eqn-saddle-extra-2}, due to Lemma \ref{lemma-DSA-Mokhtari-proof}.
First, note that,
due to identity $\mathcal{L} = I - \mathcal{W}$,
we have that \eqref{eqn-proposed} and \eqref{eqn-saddle-extra}
are the same. Next, from \eqref{eqn-proposed},
we have that:
\[
\nabla F(x^{(k)})+u^{(k)} = -\frac{1}{\alpha}x^{(k+1)} + \frac{1}{\alpha}\mathcal{W}\,x^{(k)}.
\]
Substituting this into \eqref{eqn-proposed-2},
and using $\mathcal{B}=\frac{1}{\alpha}\mathcal{W}$,
we recover~\eqref{eqn-saddle-extra-2}. Hence, the
equivalence of \eqref{eqn-proposed}--\eqref{eqn-proposed-2} and \eqref{eqn-extra}--\eqref{eqn-extra-222}
 with $\mathcal{B}=\frac{1}{\alpha}\mathcal{W}$. Next, we prove part~(a). That is, we consider \eqref{eqn-harnessing} and
\eqref{eqn-proposed}--\eqref{eqn-proposed-2}
with $\mathcal{B}=0$.
 The key is to note that $s^{(k)}$, $k=0,1,...,$
 can be written as:
 $
 s^{(k)} = u^{(k)} + \nabla F(x^{(k)}),
 $
 where $u^{(k)}$
is defined by the following recursion:
\begin{eqnarray}
u^{(k+1)} &=& \mathcal{W}\,u^{(k)} + ( \mathcal{W}-I)\,\nabla F(x^{(k)}) \nonumber \\
&=&
\label{eqn-proof-appendix-1}
u^{(k)} - \mathcal{L}\,(u^{(k)}+\nabla F(x^{(k)})),
\end{eqnarray}
for $k=0,1,...,$ and $u^{(0)}=0.$ Substituting~\eqref{eqn-proof-appendix-1} into
\eqref{eqn-proposed}, yields the desired equivalence.

\subsection{Proof of Lemma~\ref{lemma-recursion}}
Consider \eqref{eqn-proposed}. Subtracting
$x^\bullet$ from both sides of the equality,
 noting that $\mathcal{W}\,x^\bullet = x^\bullet$,
and adding and subtracting $\nabla F(x^\bullet)$ to
the term in the parenthesis on the right hand side of the equality, we obtain:
{{\small{
{\allowdisplaybreaks{
\begin{eqnarray}
e_x^{(k+1)}
 &=&
 \mathcal{W}\,e_x^{(k)}
 -\alpha \,\left( \nabla F(x^{(k)}) - \nabla F(x^\bullet) 
 + \nabla F(x^\bullet)+u^{(k)}\right) \nonumber
\\
\label{eqn-proof-append-L2}
&=&
\mathcal{W}\,e_x^{(k)}
- \alpha \,\mathcal{H}_k\,e_x^{(k)} - \alpha \,e_u^{(k)},
\end{eqnarray}}}}}
where~\eqref{eqn-proof-append-L2}
follows by the definition of $\mathcal{H}_k$,
$\mathcal{H}_k =\int_{t=0}^1 \nabla^2 F\left(x^\bullet + t \,(x^{(k)}-x^\bullet)\right)\,dt$,
 and by the definition of $e_u^{(k)}$.
 Next, consider~\eqref{eqn-proposed-2}. Using identity $\mathcal{L} = I-\mathcal{W}$, the
 equality can be equivalently written as:
 \begin{equation}
 \label{eqn-equiv-prop-2}
 u^{(k+1)} = \mathcal{W}\,u^{(k)}
 - \mathcal{L}\left( \nabla F(x^{(k)})- \mathcal{B}\,x^{(k)}\right).
 \end{equation}
Next, add $\nabla F(x^{\bullet})$ to
both sides of the equality,
and express the quantity on the right
hand side as
$\nabla F(x^{\bullet})
= \mathcal{W}\,\nabla F(x^{\bullet}) + \mathcal{L}\,\nabla F(x^{\bullet})$.
 We obtain:
 {\allowdisplaybreaks{
 \begin{eqnarray}
 e_u^{(k+1)}
 & = &
  \mathcal{W}\,(u^{(k)}+\nabla F(x^\bullet))
  - \mathcal{L}\,\left( \nabla F(x^{(k)})\right. \nonumber\\
  &-& \left.\nabla F(x^{\bullet})\right)
  + \mathcal{L}\,\mathcal{B}\,x^{(k)} \nonumber \\
   \label{eqn-to-subtract-100}
   &=&
  \mathcal{W}\,e_u^{(k)}
  - \mathcal{L}\,\mathcal{H}_k\,e_x^{(k)} + \mathcal{L}\,\mathcal{B}\,x^{(k)}.
 \end{eqnarray}}}
 Next, by the property
 $\mathcal{B} \,x^\bullet = c\,x^{\bullet}$, for some $c \in \mathbb R$,
 it follows that:
$
 \mathcal{L}\,\mathcal{B}\,x^{\bullet}
 $
 $=
 c\,\mathcal{L}\,x^{\bullet}
 $ $ = c\,[\,(I-W)\otimes I\,]\,[\mathbf{1}\otimes x^{\star}]
 $
 $
 =
 c\,[\,(I-W)\mathbf{1}\,] \otimes [\,I \otimes x^\star\,]=0.
$
 Hence,
 we can subtract
 $\mathcal{L}\,\mathcal{B}\,x^{\bullet}=0$
 from the right hand side of \eqref{eqn-to-subtract-100}
 to obtain the following:
 \begin{eqnarray}
 e_u^{(k+1)}
  &=&
  \mathcal{W}\,e_u^{(k)}
  - \mathcal{L}\,\mathcal{H}_k\,e_x^{(k)}
  + \mathcal{L}\,\mathcal{B}\,e_x^{(k)}\nonumber \\
  &=&
   \label{eqn-append-new-10}
 \mathcal{W}\,e_u^{(k)}
  + \mathcal{L}\,(\mathcal{B}-\mathcal{H}_k)\,e_x^{(k)}.
 \end{eqnarray}
Finally, note from~\eqref{eqn-append-new-10}
 that:
 \begin{equation*}
 \mathcal{J}\,
 e_u^{(k+1)} = \mathcal{J}\,e_u^{(k)}, \,\,k=0,1,...,
 \end{equation*}
 because $\mathcal{J}\,\mathcal{L}
  = [\,J \otimes I\,]\,[\,(I-W)\otimes I\,]$ $
   = [\,J (I-W)\,]\otimes I = [J-J]\otimes I=0$.
Note that
 $\mathcal{J}e_u^{(0)}
  = \mathcal{J}(0+\nabla F(x^\bullet))
   = \frac{1}{N}\mathbf{1}(\,\sum_{i=1}^N \nabla f_i(x^\star)\,)=0$.
   Thus, we conclude
   that $\mathcal{J}\,e_u^{(k)}=0$,
   for all $k$.
   Applying the latter fact to \eqref{eqn-append-new-10},
   we obtain:
\begin{equation}
\label{eqn-append-result-2}
 e_u^{(k+1)} = (\mathcal{W}-\mathcal{J})\,e_u^{(k)}
  + \mathcal{L}\,(\mathcal{B}-\mathcal{H}_k)\,e_x^{(k)}.
 \end{equation}
The relations~\eqref{eqn-proof-append-L2} and~\eqref{eqn-append-result-2}
 yield the claim of the Lemma.

\subsection{Derivation of the solution to~\eqref{eqn-21-block}
and of an approximate solution to~\eqref{eqn-21-block-cal-W}}
We first consider~\eqref{eqn-21-block}.
Note that, for any
 $\mathcal{H} \in \mathbb{H}$,
 we have that:
 $
 \|b\,I-\mathcal{H}\|=$
  $\max_{i=1,...,Nd}|b-h_i| $
  $=
 \max \left\{ |b-h_{Nd}|,\,|h_1-b|\right\}$
 $\leq
 \max\left\{|b-\mu|,\,|L-b|\right\}.$
 Here,
 $h_i$ denotes
 the $i$-th largest
 eigenvalue of~$\mathcal{H}$.
 In the last inequality
 above, we used
 the fact that
 $\mu\,I \preceq \mathcal{H}\preceq L\,I$,
 for all $\mathcal{H} \in \mathbb{H}$.
 Therefore,
 we have that:
 \begin{equation}
 \label{eqn-max-b-new}
 \max_{\mathcal{H} \in \mathbb{H}}
 \|b\,I-\mathcal{H}\|
  = \max\{|b-\mu|,\,|L-b|\}.
 \end{equation}
 The maximum in \eqref{eqn-max-b-new} is attained,
  e.g., for
  $\mathcal{H} = \mathrm{Diag}(L,\mu,...,\mu)$,
  where $\mu$ is
  repeated $(Nd-1)$ times.
 The quantity \eqref{eqn-max-b-new} is
 clearly minimized over $b \geq 0$ at
 $b = b^\star=\frac{L+\mu}{2}$.

Now, consider \eqref{eqn-21-block-cal-W},
and assume that $\lambda_N>0$.
Note that the maximal eigenvalue of
$\mathcal{W} = W \otimes I$ equals one,
and the minimal eigenvalue of
$\mathcal{W}$
equals $\lambda_N$.
 We have:
  \begin{eqnarray}
 \|b^{\prime}\,\mathcal{W}-\mathcal{H}\|
 \leq
 \label{eqn-upper-heur-new}
 \max\left\{|b^{\prime}-\mu|,\,|L-b^{\prime}\,\lambda_N|\right\}.
 \end{eqnarray}
We choose a sub-optimal
$b^{\prime}$ that
minimizes the upper bound in~\eqref{eqn-upper-heur-new}
on the desired function
$\sup_{\mathcal{H} \in \mathbb{H}}\|b^\prime\,\mathcal{W}-\mathcal{H}\|$.
It is easy to see that
the corresponding value is
$b^{\prime}=\frac{L+\mu}{1+\lambda_N}$.

\subsection{Proof of equivalence of \eqref{eqn-proposed}--\eqref{eqn-proposed-2}
and \eqref{eqn-saddle-point-abstract-prop}--\eqref{eqn-saddle-point-abstract-prop-2}}
Consider \eqref{eqn-proposed}.
From the equation, we have:
 $
\nabla F(x^{(k)})+u^{(k)} = -\frac{1}{\alpha}\left( x^{(k+1)} - \mathcal{W}\,x^{(k)}\right).
 $
Substituting the latter
relation in \eqref{eqn-proposed-2},
and using the fact
that matrices $\mathcal{L}$
 and $\mathcal{W}$ commute,
 as well as
 that $\mathcal{L}$ and
 $\mathcal{B}$ commute,
 \eqref{eqn-proposed-2}
 leads to~(25).
 Now, consider \eqref{eqn-saddle-point-abstract-prop}.
 Proceeding by the same steps as
 in deriving \eqref{eqn-saddle-extra} from \eqref{eqn-saddle-point-abstract},
 it is straightforward to verify
 that \eqref{eqn-saddle-point-abstract-prop} leads to~(24).
  Thus, the equivalence between~\eqref{eqn-proposed}--\eqref{eqn-proposed-2}
  and \eqref{eqn-saddle-point-abstract-prop}--\eqref{eqn-saddle-point-abstract-prop-2}.
  The respective equivalence for
  the method in~\cite{Harnessing}
  follows by setting $\mathcal{B}=0$ in
   \eqref{eqn-saddle-point-abstract-prop}--(29).

\subsection{Proof of Theorem~\ref{theorem-R-linear-rate} for generic
matrices~$\mathcal{B}$}

We show here that, when $\mathcal{B}=b\,I$ is replaced
with a generic symmetric matrix $\mathcal{B}$
that respects the sparsity pattern of graph~$\mathcal{G}$
 and obeys that for any $y \in {\mathbb R}^d$, there exists some $c \in {\mathbb R}$,
     such that $\mathcal{B}\,(\mathbf{1}\otimes y) = c\,(\mathbf{1}\otimes y)$, Theorem~\ref{theorem-R-linear-rate}
  continues to hold
  with $L^\prime$ replaced with
  constant $\left( L + \|\mathcal{B}\| \right)$.
 Namely, it is easy to see that
Lemma~8 continues to hold unchanged.
Further, Lemmas 9--11 pertain to
the update equation~\eqref{eqn-proposed}
that does not depend on
$\mathcal{B}$, and hence they also hold unchanged.
 The only modifications
 occur with Lemma~12. Namely,
 \eqref{eqn-proposed-2-new-3} becomes:
$
\widetilde{u}^{(k+1)}
=$
 $\widetilde{\mathcal{W}}\,\widetilde{u}^{(k)} - \mathcal{L}\,(\, \nabla F(x^{(k)}) - \nabla F(x^{\bullet})\,)
$ $+
\mathcal{L}\, \mathcal{B}\,(x^{(k)}-x^\bullet).$
Using Lipschitz continuity of $\nabla F$ and
the fact that $\|\mathcal{L}\|\leq 2$,
the latter equality implies:
 $
\|\widetilde{u}^{(k+1)}\|
$ $\leq
{\sigma}\,\|\widetilde{u}^{(k)}\| + 2\,L\,\|x^{(k)}-x^\bullet\| $ $
+ 2\|\mathcal{B}\|\,\|x^{(k)}-x^\bullet\|.
 $
 The proof of the modified Lemma~12 then proceeds
 in the same way as the remaining
 part of the proof of Lemma~12. Finally,
 the proof of the modified Theorem~4
 then also proceeds in the same way
 as the proof of Theorem~4.
\vspace{-3mm}

\bibliographystyle{IEEEtran}
\bibliography{IEEEabrv,bibliographyExactFOM}

\end{document}